\documentclass[twocolumn,floatfix,aps,pre,amsmath,showpacs]{revtex4}
\usepackage{graphicx}

\begin{document}

\title{The memory kernel of the self-intermediate
scattering function on a MD-simulated glass-forming
Ni$_{20}$Zr$_{80}$-system}
\author{ A.B. Mutiara \thanks{e-mail: amutiara@staff.gunadarma.ac.id}}

\affiliation{ Graduate Program In Information System, \\
Gunadarma University,
\\Jln. Margonda Raya 100, Depok 16424, Indonesian.}

\date{\today}

\begin{abstract}
The memory kernel of the self-part of the
intermediate scattering function is studied. We found that
the short-time behavior of the memory kernel for
our MD-simulated glass-forming Ni$_{20}$Zr$_{80}$-system shows a deviation
from the polynomial form predicted by MCT. By using the gaussian approximation
we give a more detailed description of the short-time dynamics of the system.
\end{abstract}
\maketitle
\section{INTRODUCTION}

At present the phenomena behind the liquid-glass transition and
the nature of the glassy state are not fully understood,
despite the progresses of recent years. In contrast to
usual phase transformations the glass transition seems
to be primarily dynamic in origin and therefore new
theoretical approaches have to be developed for its
description. Several theoretical models have been proposed
to explain the transition and the corresponding experimental data.
The latter concern of both \textit{the temperature dependence of particular properties},
such as the shear viscosity and the structural relaxation time,
and \textit{the time dependence repsonse of the spectra}
as visible in the dielectric susceptibility,
inelastic neutron scattering, and light scattering spectra investigations.
The spectral measurements have been extended to cover
the large frequency range from below the primary $\alpha $-relaxation
peak up to the high-frequency region of microscopic dynamics
dominated by vibrational modes \cite{Cum97}.

One of the promising approaches in this field is mode coupling theory (MCT).
MCT originally was developed to model critical phenomena \cite{Kad,Kawa}.
The non-classical behavior of the transport properties near the critical
point was thought to be caused by nonlinear couplings between slow
(hydrodynamics and order parameter) modes of the system. In later years,
the MCT was found to be applicable more generally,
to describe nonlinear effects in dense liquids \cite{Ernst} and
nonhydrodynamic effects in the case of the glass transition.

The idealized schematic model of the MCT for the liquid-glass transition,
firstly proposed by Bengtzelius et.al. \cite{Bengt84} and
independently by Leutheusser \cite{Leuth84}, implies that the liquid
(''ergodic'') state is characterized through a decay of structural
correlations while below a critical temperature $T_{c}$ a solidlike
(''nonergodic'') state exists with structural arrest
where fluctuations are effectively frozen.
In this state correlations decay on a longer time scale by
thermally activated diffusion processes \cite{Das,Gotze,Sjorgen}.

According to the MCT the time dependence of the correlation functions
for structural fluctuations follows from a damped harmonic oscillator
equation with a retarted damping term. All the details concerning
the time evolution of the structural correlations are hidden
in the memory kernel that describes the retarted damping.

In a previous paper \cite{Tei96E}, hereafter refered to as I,
one of us has evaluated this memory kernel from the time
dependence of the correlation functions in the molecular
dynamics(MD) simulations of the glass-forming Ni$_{50}$Zr$_{50}$-system.
Here two results from I shall be recalled:
first, it found that the short-time behavior of the memory kernel of
the self-part of the
intermediate scattering function $\Phi(q,t)$ shows a deviation from the
polynomial form commonly assumed in the idealized(or extended)
schematic model of the MCT. This deviation, in particular,
explains the absence of the inverse powerlaw decay of
the correlation predicted by the MCT. The deviations
originate from the vibrational motions of the atoms in
local potential minima and are restricted to the short times,
below some 0.3 ps. This, on the other hand, explains
that the MCT predictions for the late $\beta$-regime and
for the $\alpha$-peak are in a good agreement
with the experimental results \cite{Mez,Kieb,Dian,Cum92,Meg} and
MD-simulation \cite{Kob,Lew}.
Second, there is a characteristic quantity $g_m$,
as introduced in \cite{Tei96L}, that can be used
as a relative measure to describe how close
a liquid structure has approached the arrested nonergodic state.
If $g_{m}>1$ for a given temperature $T$, then
the system is able to undergo the transition
into an idealized nonergodic state with a structural arrest;
if $g_{m}<1$, the system remains in the (undercooled) liquid state ($T>T_{c}$),
and if $g_{m}\approx1 $,
the system stands in a critical situation.

The behavior of the memory kernel $F(t)$ and
the parameter $g_m$ are formally related through
the characteristic function $g(\Phi)=F(t)(1/\Phi(q,t)-1)$
introduced in \cite{Tei96L}. $g_m$ means
the maximum of $g(\Phi)$ for $\Phi\in[0,1]$ and
the short time behavior of $F(t)$ is, in particular,
visible in the behavior of $g(\Phi)$ for values $\Phi(q,t) \rightarrow 1$.
Therefore we here are interested in a detailed
description of $g(\Phi)$ for short times, respectively
$\Phi(q,t)\rightarrow 1$. Since, according to I, $\Phi(q,t)$
for short times is determined by
the vibration spectrum of the amorphous structure,
the memory kernel $F(t)$ for short times and
the corresponding $g$ also depend on the vibration spectrum.
The latter is visible in the velocity autocorrelation function (VACF)
and thus we in the following look for
an interrelationship between the VACF and $g$.
Moreover, we here consider the Ni$_{1-x}$Zr$_{x}$ system at $x=0.8$,
while in I the concentration $x=0.5$ was investigated.
Thus we ask to which extend the results for $x = 0.8$
confirm the results for $x=0.5$ from I.

Our paper is organized as follows:
In Section \ref{SIM}, we present the model and
give some details of the computations.
Section \ref{THEO} gives  a brief discussion of
some aspects of the mode coupling theory as used here and
it describes the gaussian approximation to relate $\Phi(q,t)$ and
the VACF at the short time.
Results of our MD-simulations and their analysis are
then presented and discussed in Section \ref{RD}.

\section{SIMULATIONS \label{SIM}}

As in paper I, the present simulations are
carried out as state-of-the-art isothermal-isobaric
($N,T,p$) calculations. The Newtonian equations of
$N=$ 648 atoms (130 Ni and 518 Zr) are numerically integrated
by a fifth order predictor-corrector algorithm
with time step $\Delta t$ = 2.5x10$^{-15}$s
in a cubic volume with periodic boundary conditions
and variable box length L. With regard to
the electron theoretical description of
the interatomic potentials in transition metal alloys
by Hausleitner and Hafner \cite{Haus}
we model the interatomic couplings as in \cite{Tei92}
by a volume dependent electron-gas term $E_{vol}(V)$
and pair potentials $\phi(r)$ adapted to
the equilibrium distance, depth, width, and zero of
the Hausleitner-Hafner potentials \cite{Haus}
for Ni$_{20}$Zr$_{80}$ \cite{MUT}.
For this model simulations were started through
heating a starting configuration up to 2000 K,
yielding a homogeneous liquid state.
The system then is cooled continuously to various annealing
temperatures with cooling rate $-\partial T$ = 1.5x10$^{12}$ K/s.
Afterwards the obtained configurations of
various annealing temperatures (here 1500-800 K) are
relaxed by modelling an additional isothermal annealing
that depend on the object temperature.
Finally the time evolution of these relaxed configurations
is modelled and analyzed.
Full details of the simulations are given in \cite{MUT}.

\section{THEORY \label{THEO}}

\subsection{Mode Coupling Theory \label{MoCT}}

In the MCT equation of motion for the correlation function
of structural fluctuations with wave vector
$\mathbf{q}$ can be expressed as a damped harmonic
oscillator equation (see, e.g, \cite{Gotze85,Gotze88})
\begin{equation}
\ddot{\Phi}(q,t)/\Omega _{0}^{2}+\Phi(q,t)+\int_{0}^{t}d\tau F(q,t-\tau )\dot{\Phi}(q,\tau )=0
\label{3.1}
\end{equation}
\noindent with initial conditions $\Phi(q,0)=1$ and $\dot\Phi(q,0)=0$%
. $\Omega _{0}$ is a microscopic (phonon) frequency and
$F(t)$ is the memory kernel. The schematic modell assumes
that the time dependence of $F(q,t)$
can be expressed by that of $\Phi_q(t)$ and
the idealized version of MCT, neglecting atomic diffusion, relies on
the assumption
\begin{equation}
F(q,t)=F_{0}(t):=h(t)+f(\Phi(q,t))
\label{3.2}
\end{equation}
\noindent where $f(\Phi )$ means the polynomial in $\Phi $ and $h(t)$
a short time
viscous damping which conveniently is approximated
by an instantaneous term $%
\eta _{0}\delta (t)$. The asymptotic behavior of $\Phi(q,t)$
is determined by $f(\Phi )$.

As mentioned above, there is for a given temperature a characteristic
function $g(\Phi )$ whose maximum value can be used to determine whether
the system is able to undergo a structural arrest ( $g_{m}>1$) or whether it
remains in the (undercooled) liquid state ( $g_{m}<1$ ).
This characteristic function can be derived as follows:
Given $\lim\limits_{t\rightarrow \infty }\Phi \approx \Phi ^{\infty }>0$.
Let us assume that
there is a critical time $t_{c}$ so that for $t>t_{c}$
\begin{equation}
\dot{\Phi}(t)=0 \:,
\label{3.3}
\end{equation}
\begin{equation}
\ddot{\Phi}(t)=0 \:.
\label{3.4}
\end{equation}

\noindent Substitute the equations (\ref{3.3}) and (\ref{3.4})
in MCT-Eq. (\ref{3.1}), then the MCT-Eq. becomes
\begin{equation}
\Phi ^{\infty }+f(\Phi ^{\infty })(\Phi ^{\infty }-1)=0,
\label{3.5}
\end{equation}
and after a little algebra we find
\begin{equation}
g(\Phi )=f(\Phi )(1/\Phi -1)  \label{3.6.0}
\end{equation}
\noindent with $\Phi \in \left[ 0,1\right] $.
This function is related in simple way
to $\Delta F(\Phi )=f(\Phi )-\Phi /(1-\Phi )$ frequently used in the
schematic MCT \cite{GoHaus}.

According to the extended version of MCT which simulates
atomic diffusion by taking into account the coupling to transverse currents,
the memory kernel has the following form
\begin{equation}\label{3.7}
F(t)=\mathcal{L}^{-1} \{1/[D(z)+\frac{1}{\mathcal{L}
({F_{0}(t)})_{z}}]\}
\end{equation}
Here $\mathcal{L}$ means the Laplace transform, $\mathcal{L}^{-1}$
its inverse. $D(z)$ models the coupling to the transverse currents.
$F(t)$ from Eq.(\ref{3.7}) leads to the final decay of structural
fluctuations also below $T_{c}$. This sets one of the basic problems
in classification of the solutions of Eq.(\ref{3.1}) as it may not
be obvious for a given solution whether it belongs to the regime
above or below $T_{c}$ if atomic diffusion is included.

For our analysis we introduce the Laplace Transform for the memory kernel $%
F(t)$ and the correlation function $\Phi (t)$, by taking $z=\varepsilon
-i\omega $ with $\varepsilon \rightarrow 0$, as follows
\begin{eqnarray}
  \lim_{\varepsilon\rightarrow 0}\mathcal{L}\{F(t)\}_{\varepsilon -i\omega }&=& F_{c}(\omega )+iF_{s}(\omega )
  \label{3.8.a} \\
  \lim_{\varepsilon\rightarrow 0}\mathcal{L} \{ \Phi (t)\}_{\varepsilon -i\omega }&=& \Phi _{c}(\omega )+i\Phi _{s}(\omega )
  \label{3.8.b}
\end{eqnarray}
\noindent where $\Phi _{c}(\omega )$ and $\Phi _{s}(\omega )$ are
the cosines-part and sinus-part of the Fourier transformed $\Phi
(t)$. It should be noted that according to the
fluctuation-dissipation theorem the loss part of the dynamic
susceptibility spectrum $\chi ^{\prime \prime }(\omega )$ is related
to the dynamics structure factor $\Phi _{c}(\omega )$ via

\begin{equation}
\chi ^{\prime \prime }(\omega )=\omega \Phi _{c}(\omega )
\label{3.10.a}
\end{equation}

\noindent Substitute equations (\ref{3.8.a}) and (\ref{3.8.b}) in MCT-Eq.
(\ref{3.1}),
then we obtain the equation for the spectral distribution $\omega
F_{c}(\omega )$ of the memory kernel
\begin{equation}
\omega F_{c}(\omega )=\frac{\omega \Phi _{c}(\omega )}{\left[ 1-\omega \Phi
_{s}(\omega )\right] ^{2}+\left[ \omega \Phi _{c}(\omega )\right] ^{2}} \:.
\label{3.10}
\end{equation}
Taking the inverse-Fourier transformed $F_{c}(\omega )$, we have the time
dependent memory kernel
\begin{equation}
F(t)=\frac{2}{\pi }\int_{0}^{\infty }d\omega F_{c}(\omega )\cos (\omega t) \:.
\label{3.11}
\end{equation}

\subsection{Gaussian Approximation \label{GS}}

According to the theoretical models studied in the theory of liquids
\cite{Boon,Hansen,Balucani}
we want to illustrate the correlation functions in the short-time regime for
temperatures near the critical temperature. One of the models based on a
cummulant expansion \cite{NiRah} express the correlation function in
an isotropic system
\begin{equation}
\Phi (q,t)=\exp \left\{ -q^{2}\rho _{1}(t)+q^{4}\rho _{2}(t)-\ldots \right\},
\label{3.15.0}
\end{equation}
where
\begin{eqnarray}
\rho_{1}(t)
&=&\frac{1}{6}\left\langle\left|\textbf{{x}}_{i}(t+t_{0})
{-\textbf{x}}_{i}(t_{0})\right|^{2}\right\rangle \:,
\label{3.15.a} \\
\rho _{2}(t) &=&\frac{1}{2}\left[ \rho _{1}(t)\right] ^{2}\left[
R_{2}(t)-1\right] \:,
\label{3.15.b} \\
R_{2}(t) &=&\frac{3}{5}\frac{\left\langle \left|\textbf{{x}}_{i}(t+t_{0})-%
\textbf{{x}}_{i}(t_{0})\right| ^{4}\right\rangle }{\left[
\left\langle \left|
\textbf{{x}}_{i}(t+t_{0})-\textbf{{x}}_{i}(t_{0})\right|
^{2}\right\rangle \right]^{2}} \label{3.15.c}
\end{eqnarray}
The assumption that $\Phi (q,t)$ can be expressed only by the leading term
of expansion \cite{Boon,NiRah}
\begin{equation}
\Phi (q,t)=\exp \left\{ -\frac{q^{2}}{6}\left\langle \left| \mathbf{r}\mathnormal
(t+t_{0})\mathbf{-r}\mathnormal(t_{0})\right| ^{2}\right\rangle \right\}  \label{3.16.0}
\end{equation}
is called the Gaussian approximation. Also in this approximation the
correlation function is related to one-sixth the mean square displacement
(MSD) of atoms during time $t$. In section IV. we will show our calculating
results of Eq.(\ref{3.16.0}) and prove that this approximation is correct in
the short-time regime.

To show, by using this Gaussian approximation, whether our system in the
short-time regime is a purely harmonic system or not, we need a relation
between the mean squared displacement and the spectral distribution
$Z(\omega )$ of the velocity autocorrelation function $Z(t)$ and
a assumption
that in purely harmonic system $Z(\omega )$ would be proportional to the
density of state (DOS) $Z_{dos}(\omega )$. One find that relation
\begin{equation}
\left\langle \left| \mathbf{r}\mathnormal(t+t_{0})\mathbf{-r}\mathnormal(t_{0})\right|
^{2}\right\rangle =6\omega _{0}^{2}\int_{0}^{\infty }d\omega \frac{(1-\cos
(\omega t))}{\omega ^{2}}Z(\omega ) \:,
\label{3.17}
\end{equation}
where
\begin{eqnarray}
Z(\omega ) &=&\int_{0}^{\infty }dt\cos (\omega t)Z(t)  \nonumber \\
&=&\int_{0}^{\infty }dt\cos (\omega t)\frac{\left\langle
{v}_{i}(t+t_{0})\cdot {v}_{i}(t_{0})\right\rangle }{\left\langle
\left| {v}_{i}(t_{0})\right| ^{2}\right\rangle } \:, \label{3.18}
\\
\omega _{0}^{2} &=&\frac{1}{3}\left\langle \left|
{v}_{i}(t_{0})\right| ^{2}\right\rangle =\frac{k_{B}T}{M}
\label{3.19}
\end{eqnarray}

Actually Ngai et.al.(see e.g. \cite{RoNg}) have developed a model,
which is called ''Coupling-Model'', in order to explain the dynamics of
glass-forming liquids. Here we briefly review that model. The model made use
of the assumption that vibration (phonon) and relaxation (diffusion) of the
molecules contribute independently to the density-density correlation
function; thus, $\Phi (q,t)$ is equal to the product $\Phi _{ph}(q,t)\times
\Phi _{relax}(q,t)$ \cite{Zorn}. In this model the correlation
function of the relaxation $\Phi _{relax}(q,t)$%
\begin{equation}
\Phi _{relax}(q,t)=\left\{
\begin{array}{ll}
\exp (-t/t_{0}(q,T)) & \mbox{ for $t<t_{c}$} \\
\exp (-(t/\tau (q,T))^{1-n}) & \mbox{ for $t>t_{c}$}
\end{array}
,\right.
\label{3.20.a}
\end{equation}

where respectively $t_{c}$ is a temperature-independent crossover
time separating two time regimes in which the relaxation dynamics
differ, $t_{0}$
the Debye-relaxation time, $\tau $ the Kohlrausch-relaxation time, and $%
1-n=\beta $ the Kohlrausch stretching exponent

The phonon contribution to $\Phi (q,t)$ is determined by the harmonic DOS of
the phonon modes, $Z_{dos}^{harm}(\omega )$, according to the formula (see
e.g.\cite{Zorn,Lov}), in a classical limit,
\begin{equation}
\Phi _{ph}(q,t)=\exp (-\frac{q^{2}}{2}u(t,T)) \:,
\label{3.20}
\end{equation}
where
\begin{equation}
u(t,T)=\frac{2k_{B}T}{M}\int_{0}^{\infty}d\omega \frac{(1-\cos
(\omega t))}{\omega ^{2}}Z_{dos}^{harm}(\omega ).  \label{3.21}
\end{equation}
$u(t,T)$ decreases with time, leveling off to constant value $%
u(T)=\lim\limits_{t\rightarrow \infty }u(t,T)$. Correspondingly, $%
\Phi (q,t)$ has the value $\exp (-q^{2}u(T)/2)$, which is the well known
Lamb-M\"{o}ssbauer factor. For a single Einstein oscilator of frequency $\omega_{e}$ the harmonic DOS of the phonon modes is given by
\begin{equation}
  Z_{dos}^{harm}(\omega)=\delta(\omega-\omega_{e})
 \label{3.21.1}
\end{equation}
and, for a Debye model of a solid,
\begin{equation}
  Z_{dos}^{harm}(\omega)=\left\{
       \begin{array}{ll}
         3\omega^2/\omega_{D}^3 & \mbox{for $\omega<\omega_{D}$}\\
         0                      & \mbox{for $\omega<\omega_{D}$}
    \end{array}
      \right.
  \label{3.21.2}
\end{equation}

By comparing this expressions of the coupling model with those of the theory
of liquids and making a assumption that $Z_{dos}^{harm}(\omega )$ and $%
Z_{dos}(\omega )$ are the same quantities in the short-time regime we find
that $\Phi _{ph}(q,t)$ (Eq.(\ref{3.20}) with Eq.(\ref{3.21})) of the
coupling model is the same as $\Phi (q,t)$ (Eq.(\ref{3.16.0}) with Eq.(%
\ref{3.17})) of the theory of liquids (with the Gaussian approximation).

Considering that the model of the theory of liquid has a simple form we use
for our further analysis that model. We believe that by making use of the
model it is enough to prove whether in the short-time regime our system is a
harmonic system or not.

Following Cummins et.al.\cite{Cum97} the correlation functions
$\Phi(t)$ and the corresponding susceptibility spectrum
$\chi^{\prime\prime }(\omega )$ can be considered to consist of
three regimes: 1) at high frequencies (short times) a nearly
temperature-independent microscopic structure; 2) at low
frequencies (long times) a strongly temperature-dependent
$\alpha$-relaxation peak associated with structural relaxation and
diffusion process; 3) (intermediate frequencies) between a) and b)
a minimum in $\chi^{\prime \prime }(\omega)$ (or ''plateau'') in
$\Phi(t)$ whose amplitude and position is also temperature
dependent.

From above definition of regions we write now in general form the full
susceptibility spectrum
\begin{equation}
\chi ^{\prime \prime }(T,\omega )=\chi _{st}^{\prime \prime }(T,\omega
)+\chi _{it}^{\prime \prime }(T,\omega )+\chi _{lt}^{\prime \prime
}(T,\omega )  \label{3.12}
\end{equation}
After our intensive study about the short-time regime for the object
temperatures near the critical temperature it is a little difficult to
separate purely the short time regime from the intermediate regime. Also it
is important to keep in mind that our definition for the short-time part in
Eq.(\ref{3.12}) include a small part of the $\beta $-relaxation process.

\begin{figure}[htbp]
\includegraphics[width=8.cm, height=5.cm]{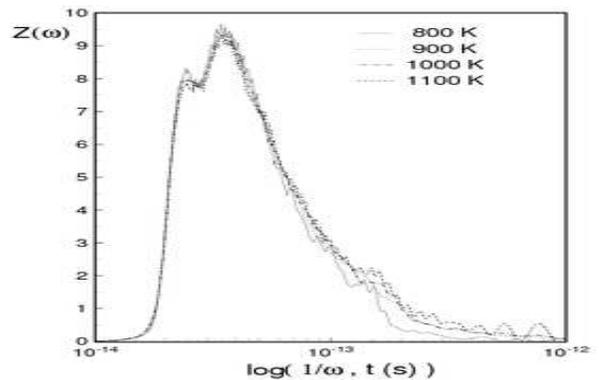}
\caption{The spectral distribution of the velocity
autocorrelation function for different temperature near $T_{c}$.}
\label{Vacfspe}
\end{figure}

Considering that the spectral distribution of the velocity autocorrelation
function $Z(t)$ for each temperature near $T_{c}$ ( from $\left[ 19\right] $
$T_{c}=1050\pm 25K$ ) shows too in our system a nearly
temperature-independent form (see Fig.~\ref{Vacfspe}), we can scale now the short-time
part of the susceptibility spectrum, so that
\begin{equation}
\chi _{sc}^{\prime \prime }(T,\omega )=\frac{T}{T_{0}}\chi _{M}^{\prime
\prime }(\omega ) \:,
\label{3.13}
\end{equation}
where $\chi _{M}^{\prime \prime }$ is the master curve of the susceptibility
spectrum that is now a temperature-independent quantity and is defined
\begin{equation}
\chi _{M}^{\prime \prime }(\omega )=\frac{1}{N}\sum_{i}^{N}\frac{T_{0}}{T_{i}
}\chi _{st}^{\prime \prime }(T,\omega ) \:,
\label{3.14}
\end{equation}
where $N$ is the number of the object temperature and $T_{0}$ is a arbitrary
normalization of temperature.


\section{RESULTS AND DISCUSSION \label{RD}}
The central object of our analysis is the self-part of the intermediate
scattering function:
\begin{equation}
\Phi (\textbf{{q}},t)=\left\langle \exp \left\{ i\textbf{{q}}\cdot
[\textbf{{x}}_{i}(t+t_{0})-\textbf{{x}}_{i}(t_{0})]\right\}
\right\rangle . \label{1}
\end{equation}
The bracket means averages over the atom $i$ and the initial configurations $t_{0}$.

In Figs.~\ref{Pmdmctq8} and~\ref{Pmdmctq6} we show the results evaluated
from our MD data (with symbols) for
wave vector $q=\frac{2\pi }{L}(n,0,0)$ with n = 8 and 6 which correspond
respectively to $\left| q\right| =19.7$ nm$^{-1}$ and $\left| q\right| =14.7$
nm$^{-1}$. Both figures present the average over the star of six $q$ vectors
which are equivalent on assuming equivalence of the simulation cube axis and
their inverses.

\subsection{MCT-Analysis \label{Mctana}}

The following Fig.~\ref{Pmdmctq8} and~\ref{Pmdmctq6} present the self-part of the intermediate
scattering function for different object temperatures according to MCT (with
line) and our MD-simulation (with symbols).
The discrepancy between both results are considerable as results from the bad
fitting of the Kohlrausch-law to our MD-data and the inaccurracy by
calculation (see \cite{Tei96L,Tei96E,MUT} for details of the algorithm used for calculation).

\begin{figure}[htbp]
 \includegraphics[width=8.cm, height=5.cm]{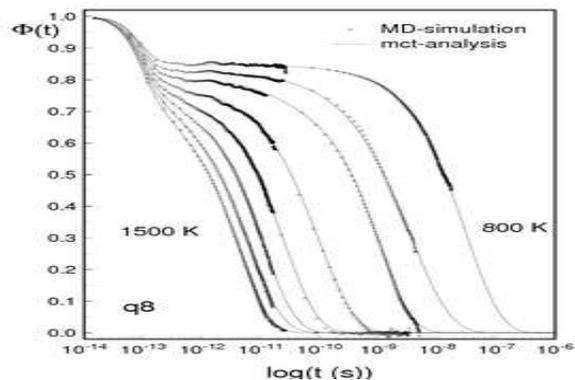}
 \caption{The self-part of the intermediate
 scattering function for wave vector $q_{8}$ from MD-data (with symbol) and from MCT-Equations (with line).
\label{Pmdmctq8}}
\end{figure}
\begin{figure}[htbp]
\includegraphics[width=8.cm, height=5.cm]{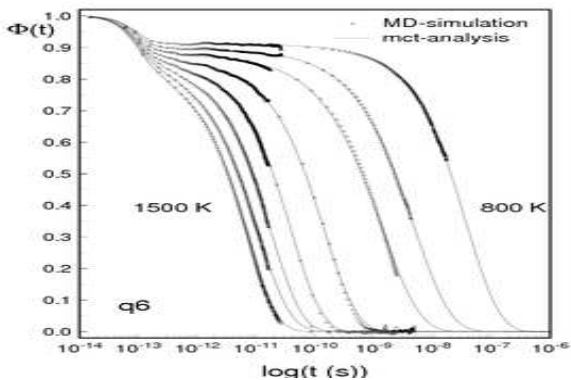}
\caption{The same as Fig.~\ref{Pmdmctq8} but for $q_{6}$.
\label{Pmdmctq6}}
\end{figure}

From both figures we found that the self-part of the intermediate
scattering function show for $q_{8}$ and $q_{6}$ at average temperatures the
structural relaxations that happen in three succesive steps. According to
predictions of the idealized schematic model of MCT the correlator should
decay in three succesive steps [20] at this temperatures. The first is a
fast initial decay on the time scale of vibrations of atoms ($t<1$ ps). This
step is characterized by MCT only global. The second is the $\beta $%
-relaxation regime (typically in the range 1 ps $<t<$1 ns). In the early $%
\beta $-relaxation regime the correlator should decrease according to $\Phi
(t)=f_{c}+A/t^{a}$ and in the late $\beta $-relaxation regime, which appear
only in melting, according to von Schweidler-law $f_{c}-Bt^{b}.$ Between
them the wide plateau appear near the critictical temperature $T_{c}$.
In the case of idealized glass the von Schweidler-law does not appear more,
but $\Phi (t)=f_{c}+A/t^{a}$ flows into the konstant asymptotic value $f_{c}$%
. In melting the $\alpha $-relaxation appear as the last step that results
from the decay after the von Schweidler-law and could be described by the
Kohlrausch-law $\Phi (t)=A_{0}\exp (-(t/\tau _{0})^{\beta })$ whose the
relaxation time $\tau _{0}$ near the glass transition shifts dramatic to the
longer time scale.

The decay for the early $\beta$-regime, which according to MCT
follows the inverse power-law $\Phi \sim f_{c}+A/t^{a}$, could not
be observed in our data. The reason for it is that in our system
$\Phi_{m}(T)$ (the maximum of g$(\Phi)$) and $\Phi_{o}(T)$ (the
threshold-value of $G(\Phi)$, see Eq.(\ref{Gphi})) are nearly
parallel, which cause that the power-law of the early
$\beta$-regime is dressed up. On the one side, for this regime the
so-called ``Boson-peak'' is discussed by several authors (see
e.g.\cite{Sok,Has,Hab}), on the other side one called this peak as
the $\beta$-maximun \cite{Mur95}. The first group of authors
claimed that as a result of the atomic vibrations in the
$\beta$-regime the boson-peak in a glass-system is observed, which
depends on the fragility of the system. The more fragile the
system ist, the weaker is the effects of the atomic vibrationen,
i.e., the height of a boson-peak will be observed smaller and not
clearer. As a result the relaxation-process is here observed
clearer. In a connection with the fragility the glass-systems can
be classified into two groups \cite{Angel,Sok}: 1) fragile
glass-systems, which have no directional bonds, e.g., Van der
Walls and ionic systems. These glass formers exhibit strong
non-Arrhenius-like increase of the viscosity upon colling; 2)
strong glass-systems, which have strong directional bonds. These
glass-systems have the temperature dependence of the viscosity
that is more Arrhenius like and weaker.

In a connection with this classification it is until today still
difficult to determine whether the metallic-glass belongs to a
strong glass-system or to a fragile glass-system, because usually
the behavior of the viscosity in a metallic glass-system lies
between both borderline case. So far from in hand analysis of our
data we can state nothing about the ``Boson-peak''.

We observed that the self-part of the intermediate scatteing
function for $T < T_{c}$ shows a small bump at $t \approx
2.5x10^{-12}$ s. Th same phenomen war observed on MD-simulations
for a OTP-system by Lewis and Wahnstr\"om \cite{Lew}and for a
binary Lenard-Jones mixture by Kob and Andersen \cite{Kob}. This
bump is interpreted by those authors as a result of the
finite-size effect. The same bump was also observed on another
Lenard-Jones mixture-system \cite{Wahn}, on a liquid-salt system
\cite{Sig}, and on a colloidal-suspenstion system \cite{Low}.

We have also analyzed that the height of plateaus as well as the
time scale, at which the self-part of the intermediate scattering
function decay finally to zero, depend strongly on wave vector
$q$. Our MD-results of $\Phi(q,t)$ for the longer time show that
$\Phi(q,t)$ always decay to zero at both $T < T_{c}$ and
$T>T_{c}$. We interpret this result within the scope of the
extended schematic MCT. The extended schematic MCT predicts that
the long-time behavior of $\Phi(q,t)$ always decay to zero, when
thermally activated hopping-processes is taken to be account in a
system. According to Teichler \cite{Tei96E,Tei96L,Tei97} and
Aspelmeier \cite{Timo95} these hopping-processes take place on a
MD-simulated Ni$_{50}$Zr$_{50}$. As one can there observe, the
reason for decaies of $\Phi(q,t)$ on the longer time scale is that
on this time scale  thermally activated hopping-processes run off.
To prove this statement, we have to investigate the
selft-diffusion konstant of atoms in our atoms. The MCT predicts
that self-diffusion konstant follows a power-law that is a
function of $T_{c}$:
\begin{equation}
  D_{\alpha} \propto (T-T_{c})^{\gamma}
  \label{4.1}
\end{equation}
where $\gamma=1/(2a)+1/(2b)$ is valid and by knowing of the
exponent $\lambda$ can be calculated. $\alpha$ is the specific
atom. The estimation of $\gamma$ ove that expression is only
possible, if a $q$-independence exponent $\lambda$ is given as
required by deriving that expression. On MD-simulation we can
determine for different temperatures of interest the
self-diffusion konstant throught the calculation of mean squared
displacement (MSD). Then we find the curve $D_{\alpha}$ vs $T$.
The effective exponent ${\gamma}$ can be determine by using the
least-squared method. From the curve we can prove whether the
thermally activated hopping-processes run off at $T<T_{c}$ or
tempertures near $T_{c}$ in our system or not, as we investigate
whether there is a deviation from (\ref{4.1}) or not. If there is
this deviation, then it means that the hopping processes still
take place at $T<T_{c}$ or $T$ near $T_{c}$ \cite{Tei97,HaYip}.

From above discussion we can state that the behavior of
$\Phi(q,t)$ in our system agree with the prediction of the
extended MCT, in which the long-time behavior of $\Phi(q,t)$
decays to zero as a result of the atomic diffusion.

The MCT predicts that the behavior of $\Phi(q,t)$ in
the last $\alpha$-regime can be good approximated
by the Kohlrausch-law. We have determined
the Kohlrausch's parameters $A_{0}$, $\beta$, and $\tau_{0}$.
Our results show that the parameter in our system
(see Table \ref{table1}) depends weakly on wave vectors $q$ and
varies with temperature.
The $\beta$-value increases with a decreasing $q$-value about 0.05.
This result is similar as one by Lewis and Wahnstr\"om
in MD-simulated OTP-system \cite{Lew}.

On our results we have noticed that the $\beta$-value seem to
increase to one at higher temperatures with a decreasing
$q$-value. This phenomane is also found in another MD-simulation
by Bernu et.al. \cite{Bernu}. The MCT predicts that $\beta$-value
for $q\sim q_{0}$ ($\sim$ 20 nm$^{-1}$), which corresponds to a
nearest-neighbor distance, is found typically in range 0.6
$<\beta<$ 0.8 that depends on the system of interest; for $q \ll
q_{0}$ the $\beta$-value seem to increase. Also our $\beta$-value
agrees well with the MCT''s predictions.

According to MCT the $\alpha$-relaxation time diverge near $T_{c}$
after the power-law \cite{GoSRP,HaYip}
\begin{equation}
  \tau_{0}(T) \propto (T-T_{c})^{\gamma}
  \label{4.2}
\end{equation}
where $\gamma$ is the same exponent in Eq.(\ref{4.1}). On our
results we have found that the behavior of the power-law was
broken in the nearest environments from $T_{c}$, because as a
result of thermally activated atomic diffusion the
$\alpha$-relaxation time $\tau_{0}$ kept to a finite-value.

Through a viewing of the $\alpha$-relaxation time $\tau_{0}$ we
have found that this relaxation time has a strong depending on
$q$-value. The same result was observed in a MD-simulated binary
Lenard-Jones mixture by Kob and Andersen \cite{Kob} and in a
experimentel quasy-elastic neutron scaterring on three polymers by
Colmenero et.al. \cite{Col}. We have noticed that there is an
anomaly on the $\alpha$-relaxation time, namely, the relaxation
time increases drastic near the glass-transition. One explains
this anomaly \cite{TeiUv95}: the system is found above $T_{c}$ on
the way to a metastable equilibrium of undercooled liquid-phases.
Then the system comes below $T_{c}$ into a unstable state, where
the relaxation processes take place in a direction of the
equilibrium. On grounds of this anomaly we can state that the
glass transition actually is a dynamic transition which is more in
the change of the art on atomic movements and less in the change
of the structure \cite{Roux,Timo95}

\begin{figure}[htbp]
\includegraphics[width=8.cm, height=5.cm]{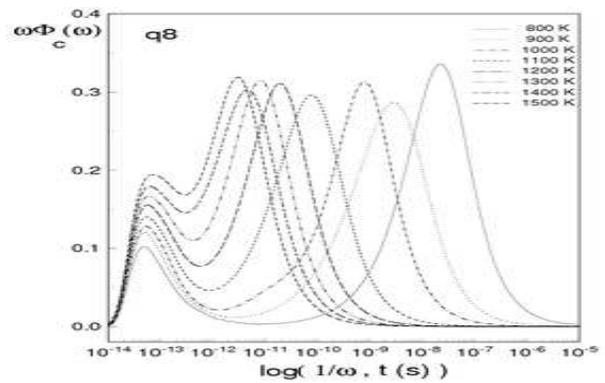}
\caption{The dynamic susceptibility for wave vector $q_{8}$
\label{Mdkq8}}
\end{figure}

\begin{figure}[htbp]
\includegraphics[width=8.cm, height=5.cm]{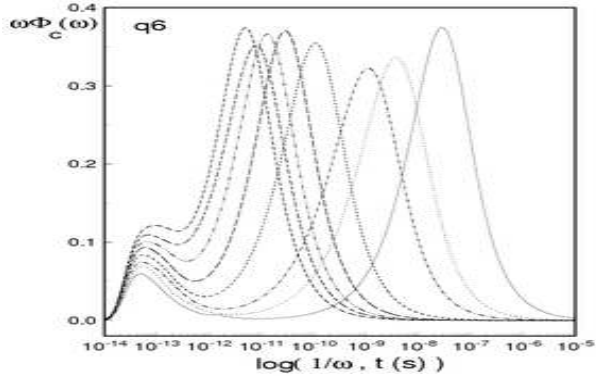}
\caption{The same as Fig. ~\ref{Mdkq8}, but for wave vector
$q_{6}$ \label{Mdkq6}}
\end{figure}

Figures~\ref{Mdkq8} and ~\ref{Mdkq6} present the dynamic
susceptibility. From both figures we found that there are three
different frequency-regimes of the dynamics susceptibility as
mentioned in theory section. As predicted by extended MCT the
$\alpha$-peak shift with a decreasing temperature to the lower
frequency, und this peak and the minimum of the dynamics
susceptibility are no more dissapear at temperature near $T_{c}$.
The extended MCT assumes this behavior, when thermally activited
hopping-processes are included formally in the memory kernel
$F(t)$. These processes has taken place in our system, i.e., these
predictions agree with statements mentioned above.

\begin{table}
\begin{center}
\caption{The Kohlrausch parameters $A_{0}$, $\tau_{0}$, $\beta$
fitted to $\Phi(q,t)$ from MD-data for wave vector $q_{8}$ and
$q_{6}$.} \label{table1}
\begin{tabular}{r|lrl|lrl}
\multicolumn{1}{c|}{} &
\multicolumn{3}{c|}{$q_{8}=19.7$ [nm$^{-1}$]} &
\multicolumn{3}{c} {$q_{6}=14.7$ [nm$^{-1}$]} \\
\hline
\multicolumn{1}{c|}{$T$ [K]} &
\multicolumn{1}{c}{$A_{0}$} &
\multicolumn{1}{c}{$\tau_{0}$[ns]} &
\multicolumn{1}{c|}{$\beta$} &
\multicolumn{1}{c}{$A_{0}$} &
\multicolumn{1}{c}{$\tau_{0}$[ns]} &
\multicolumn{1}{c}{$\beta$} \\
\hline
1500 & 0.690 & 0.004 & 0.807 & 0.821 & 0.007 & 0.917 \\
1400 & 0.729 & 0.007 & 0.820 & 0.827 & 0.011 & 0.821 \\
1300 & 0.746 & 0.012 & 0.840 & 0.835 & 0.018 & 0.911 \\
1200 & 0.759 & 0.024 & 0.780 & 0.859 & 0.042 & 0.781 \\
1100 & 0.761 & 0.095 & 0.740 & 0.860 & 0.146 & 0.795 \\
1000 & 0.790 & 0.966 & 0.748 & 0.859 & $\sim$ 1.400 & 0.813 \\
 900 & 0.826 & $\sim$ 2.448 & 0.680 & 0.894 & $\sim$ 4.871 & 0.741 \\
 800 & 0.848 & $\sim$ 31.100 & 0.756 & 0.911 & $\sim$ 38.640 & 0.794 \\
\end{tabular}
\end{center}
\end{table}

We have noticed that the height of $\alpha$-peak for each
temperature depends strongly on $q$-value, namely, the height of
$\alpha$-peak decreases with a increasing $q$-value. The
microscopic (phonon)-regime as well as the height of this regime
decreases with a decreasing both $q$-value and $T$. This result
could be simply interpreted, if one takes into consideration the
height of plateaus (the non-ergodicity parameter $f_{c}$) on the
intermediate scattering function, which depends strongly on the
$q$-value. Because the height of the $\alpha$-peak is proportional
to the height of plateaus, and the microscopic-peak is
proportional to $1-f_{c}$, then the above mentioned dependence of
the both height-peaks results directly from the $q$-dependence
non-ergodicity paramater.


\begin{figure}[htbp]
\centering \leavevmode
\includegraphics[width=8.cm, height=5.cm]{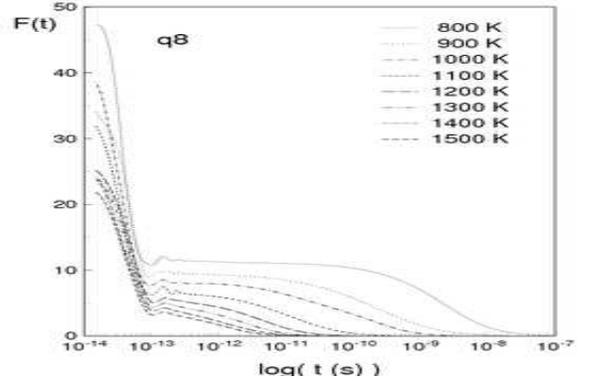}
 \caption{The kernel $F(t)$ for wave vector $q_{8}$.
\label{Ft8}}
\end{figure}

\begin{figure}[htbp]
\centering \leavevmode
\includegraphics[width=8.cm,height=5.cm]{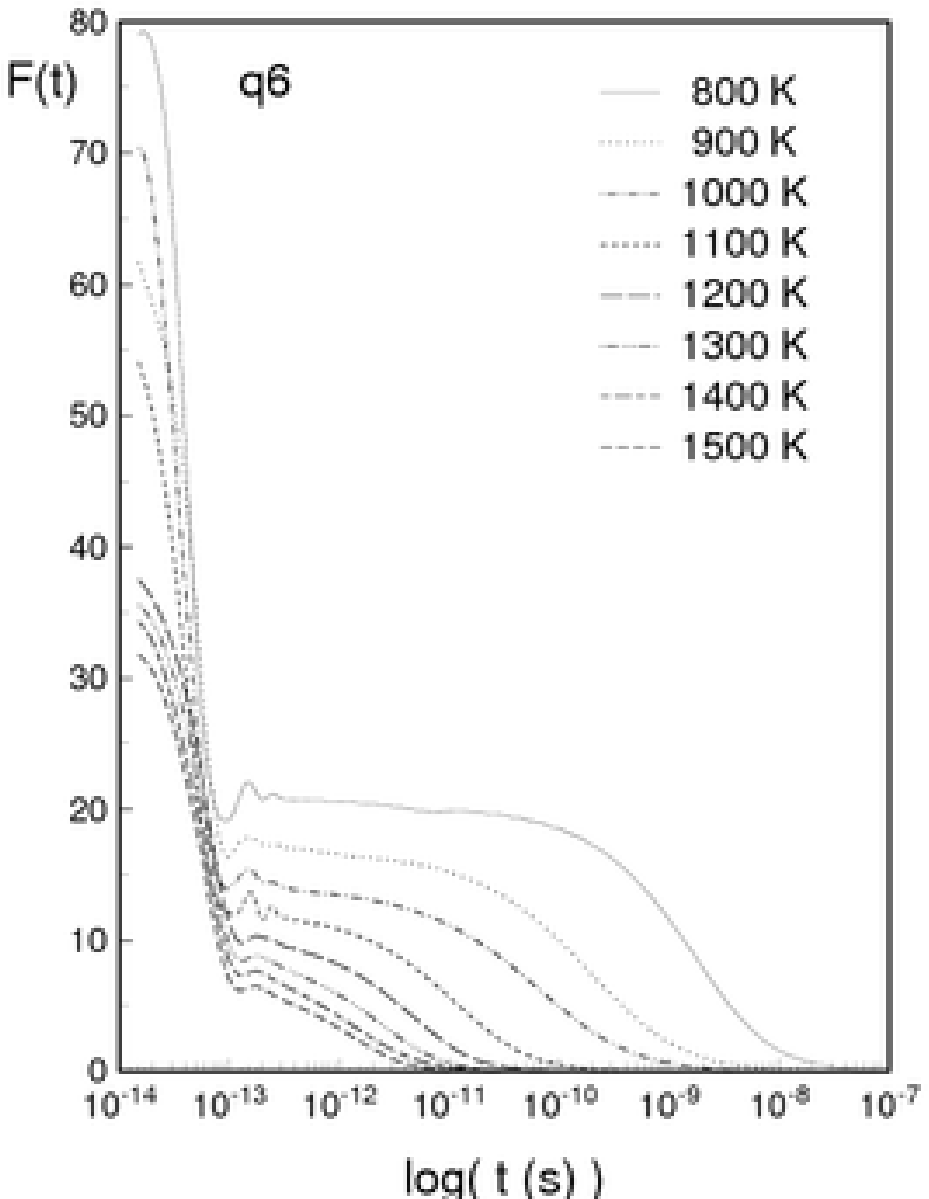}
\caption{The same as Fig.~\ref{Ft8} but for wave vector $q_{6}$.
\label{Ft6}}
\end{figure}

Figures ~\ref{Ft8} and ~\ref{Ft6} present our evaluated kernels
$F(t)$ from MD-data of $\Phi(q,t)$ for both wave vectors of
interest. From both figures we found that there is a
temperature-dependence threshold-value $\Phi_{0}(T)$ (
$\Phi_{0}(T)\in [0;1]$ ), where under $\Phi_{0}(T)$ the behavior
of $F(t)$ shows a course that corresponds to a polynom with
positive coefficients, and otherwise this behavior deviats from a
polynomial form. According to the prediction of the idealized MCT
the kernel $F(t)$ as well as $f(\Phi)$ should show a polynomial
form. Also our results do not agree with this prediction for all
of regimes, but only for the regime that lies under $\Phi_{0}(T)$.
This is understandable, because the idealized MCT does not fully
describe the atomic vibrations of the system.

We have noticed that $\Phi_{0}(T)$ depends strongly on the
$q$-value and weakly on the change of temperature. $\Phi_{0}(T)$
shift with a decreasing both $q$-value and $T$ to higher values.

\begin{figure}[htbp]
\centering \leavevmode
\includegraphics[width=8.cm, height=5.cm]{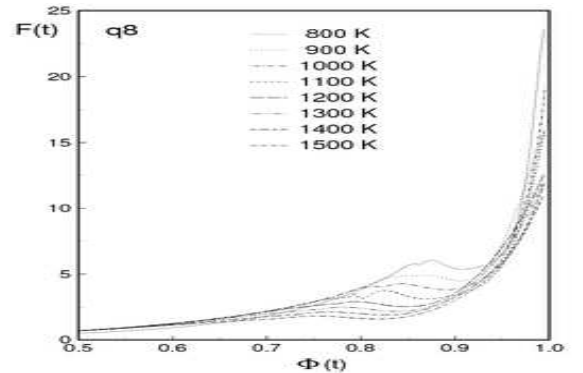}
\caption{The kernel as a function of $\Phi$ for wave vector $q_8$.
\label{Ftpt8}}
\end{figure}

\begin{figure}[htbp]
\centering \leavevmode
\includegraphics[width=8.cm, height=5.cm]{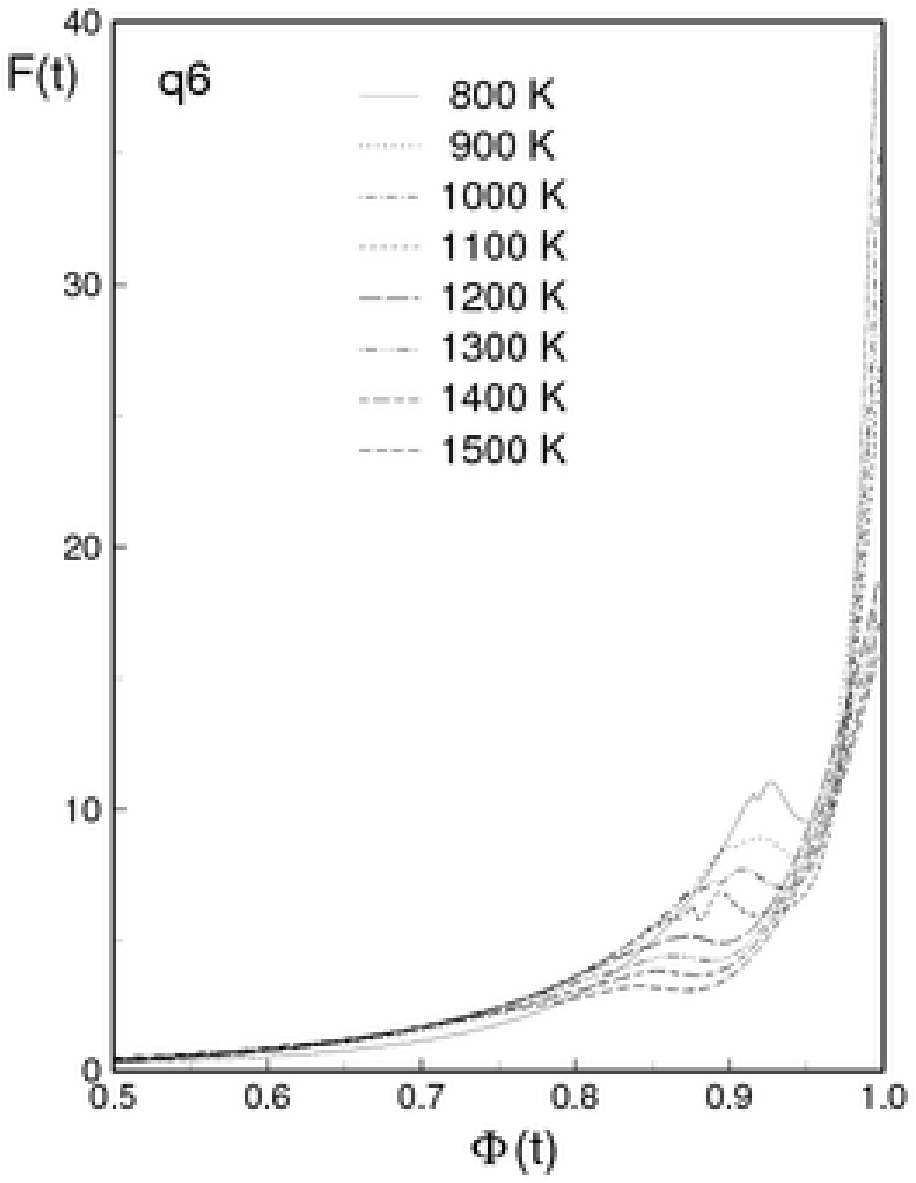}
\caption{The same as Fig.~\ref{Ftpt8} but for wave vector $q_{6}$.
\label{Ftpt6}}
\end{figure}

\begin{figure}[htbp]
\centering \leavevmode
\includegraphics[width=8.cm, height=5.cm]{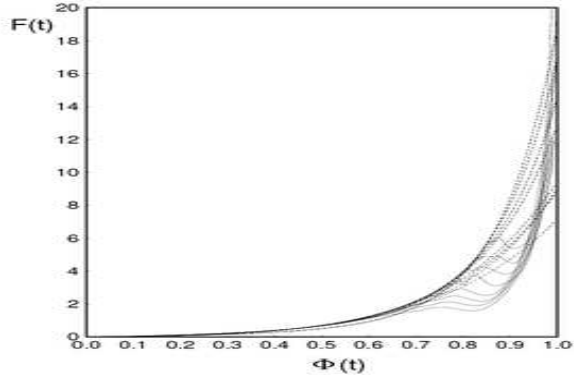}
\caption{The kernel as function of $\Phi$ for wave vector $q_8$
(dotted line: extrapolated low-$\Phi$ polynomial).
\label{Fphianp8}}
\end{figure}

\begin{figure}[htbp]
 \centering \leavevmode
\includegraphics[width=8.cm, height=5.cm]{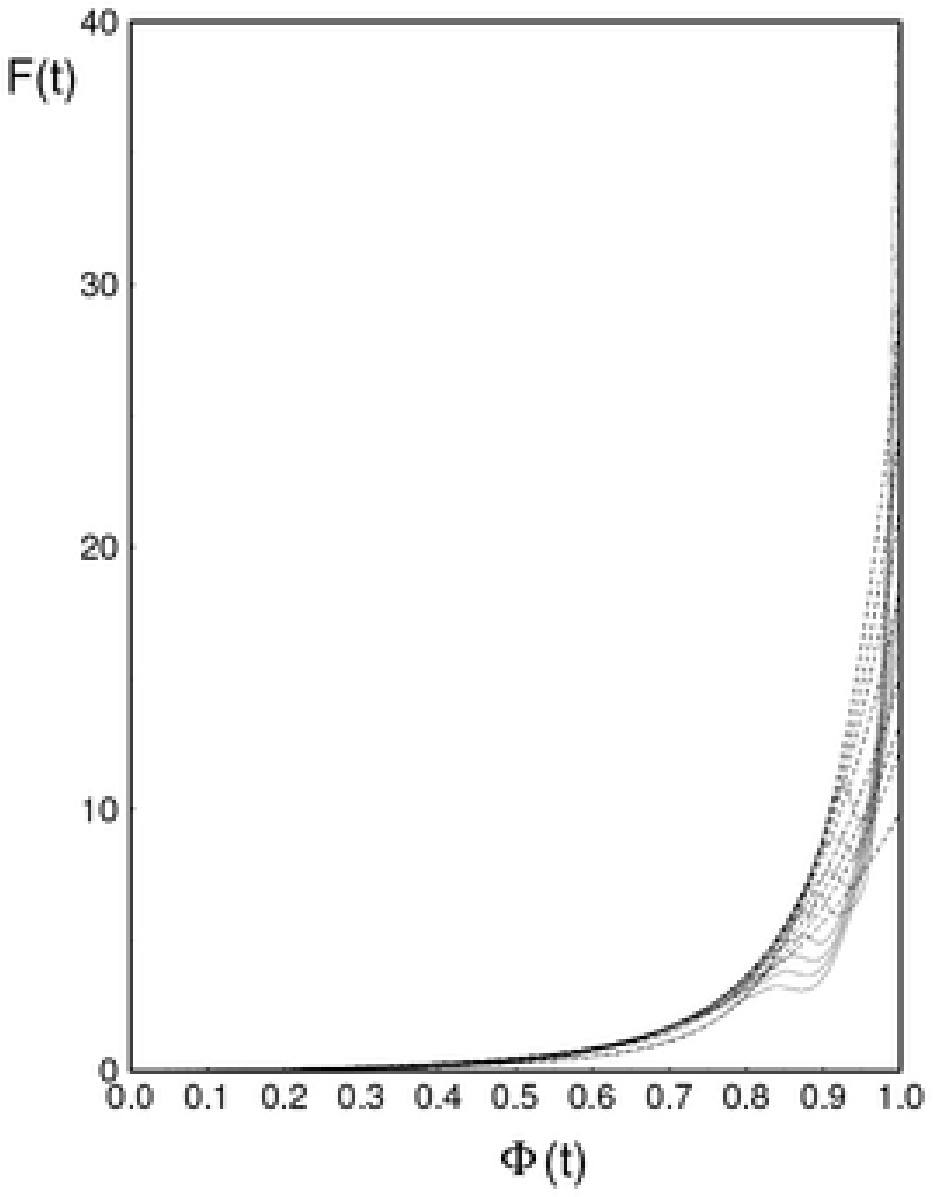}
 \caption{The same as Fig.~\ref{Fphianp8} but for wave vector $q_{6}$.
 \label{Fphianp6}}
\end{figure}

There is here a important point that we can stretch, namely, by a
calculating of the kernel $F(t)$ we can make for $\Phi(T)<
\Phi_{0}(T)$ a new model of the memory kernel $f(\Phi)$, which
presents of course a polynomial form. By a extending for
$\Phi(T)>\Phi_{0}(T)$ we can then determine the term, which could
be assigned to the atomic vibrations of the system and corresponds
to the difference between the actually kernel $F(t)$ and the
fitted memory kernel $f(\Phi)$. To this purpose we present our
calculating for the kernel $F(t)$ as a function $\Phi$ in Fig.
~\ref{Ftpt8} and ~\ref{Ftpt6}. According to MCT one define
$f(\Phi)$ as follows :
\begin{equation}
 f(\Phi)=\sum_{n} \lambda_n\Phi^n
  \label{FPhi}
\end{equation}
where $\lambda_n$ are positive coefficients. We have fitted
Eq.(\ref{FPhi}) to the kernel $F(\Phi)$. (see Fig. \ref{Fphianp8}
and \ref{Fphianp6}), and then we have for our system positive
coefficients $\lambda_n$, which are given in Tables \ref{table2}
and \ref{table3}


\begin{table}
\begin{center}
\caption{Coefficients of the polynomial expansion $f(\Phi)=\sum_n
\lambda_n \Phi^n$ for wave vector $q_8$.}
\label{table2}
\begin{tabular}{rrrrrrrrr}
\multicolumn{1}{c}{} &
\multicolumn{8}{c}{$\lambda_n$} \\
\hline
\multicolumn{1}{c}{} &
\multicolumn{8}{c}{$T$ K} \\
\hline
\multicolumn{1}{r}{$n$} & \multicolumn{1}{r}{$800$} &
\multicolumn{1}{r}{$900$}& \multicolumn{1}{r}{$1000$} &
\multicolumn{1}{r}{$1100$} & \multicolumn{1}{r}{$1200$} &
\multicolumn{1}{r}{$1300$} & \multicolumn{1}{r}{$1400$} &
\multicolumn{1}{r}{$1500$} \\
\hline
1 & 0.255 & 0.423 & 0.232 & 0.357 & 0.271 & 0.253 & 0.308 & 0.198 \\
2 & 0.714 & 0.794 & 0.658 & 0.688 & 0.661 & 0.598 & 0.820 & 0.800 \\
3 & 0.040 & 0.597 & 1.757 & 1.177 & 1.174 & 1.630 & 1.165 & 1.592 \\
4 & 1.390 & 1.318 & 2.144 & 1.621 & 2.089 & 1.850 & 1.876 & 2.628 \\
5 & 2.218 & 1.930 & 1.637 & 1.655 & 2.535 & 2.116 & 1.439 & 0.057 \\
6 & 1.586 & 1.628 & 0.700 & 1.817 & 2.414 & 1.828 & 0.593 & 1.769 \\
7 & 0.321 & 0.621 & 0.000 & 1.445 & 0.232 & 0.566 & 0.742 & 0.040 \\
8 & 0.000 & 0.000 & 0.009 & 1.251 & 0.012 & 0.000 & 2.089 & 0.185 \\
9 & 2.438 & 1.575 & 1.552 & 0.623 & 0.032 & 0.000 & 0.002 & 0.143 \\
10 & 9.864 & 8.069 & 5.907 & 2.253 & 0.040 & 0.020 & 0.000 & 0.451

\end{tabular}
\end{center}
\end{table}

\begin{table}
\begin{center}
\caption{Coefficients of the polynomial expansion $f(\Phi)=\sum_n
\lambda_n \Phi^n$ for wave vector $q_6$.}
\label{table3}
\begin{tabular}{rrrrrrrrr}
\multicolumn{1}{c}{} &
\multicolumn{8}{c}{$\lambda_n$} \\
\hline \multicolumn{1}{c}{} &
\multicolumn{8}{c}{$T$ K} \\
\hline \multicolumn{1}{r}{$n$} & \multicolumn{1}{r}{$800$} &
\multicolumn{1}{r}{$900$} & \multicolumn{1}{r}{$1000$} &
\multicolumn{1}{r}{$1100$} & \multicolumn{1}{r}{$1200$} &
\multicolumn{1}{r}{$1300$} & \multicolumn{1}{r}{$1400$} &
\multicolumn{1}{r}{$1500$} \\
\hline
1 & 0.235 & 0.316 & 0.362 & 0.246 & 0.173 & 0.189 & 0.240 & 0.162 \\
2 & 0.021 & 0.386 & 0.373 & 0.246 & 0.394 & 0.278 & 0.394 & 0.135 \\
3 & 0.918 & 0.924 & 0.990 & 0.965 & 0.214 & 0.612 & 0.682 & 0.515 \\
4 & 1.291 & 0.994 & 1.240 & 1.006 & 0.529 & 0.810 & 0.640 & 1.397 \\
5 & 0.016 & 0.588 & 0.873 & 0.127 & 1.448 & 1.323 & 1.563 & 2.250 \\
6 & 0.102 & 0.130 & 0.289 & 0.009 & 2.594 & 2.528 & 2.250 & 4.494 \\
7 & 0.001 & 0.000 & 0.000 & 1.780 & 3.493 & 2.873 & 2.756 & 0.689 \\
8 & 0.000 & 0.249 & 0.366 & 4.877 & 3.671 & 4.101 & 1.277 & 0.185 \\
9 & 0.001 & 1.238 & 1.893 & 6.649 & 2.645 & 0.563 & 1.166 & 0.063 \\
10 & 0.000 & 3.152 & 5.013 & 2.365 & 0.020 & 0.003 & 0.003 & 0.001 \\
11 & 3.192 & 7.167 & 10.157 & 0.092 & 0.030 & 0.303 & 1.323 & 0.003 \\
12 & 8.222 & 7.978 & 0.002 & 0.044 & 0.042 & 0.005 & 0.001 & 0.014 \\
13 & 15.148 & 1.400 & 0.000 & 0.013 & 0.011 & 0.002 & 0.002 &
0.000
\end{tabular}
\end{center}
\end{table}

To calculate Eq.(\ref{3.6.0}), as in I, we introduce the following
equations
\begin{equation}
   G(\Phi):= F(\Phi)(1/\Phi-1)
 \label{Gphi}
\end{equation}
\begin{equation}
  g(\Phi)= P(\Phi)(1/\Phi-1)
  \label{gphi}
\end{equation}
Eq.(\ref{gphi}) is analog to Eq.(\ref{3.6.0}), where $f(\Phi)$ in
$g(\Phi)$ is substituted by $P(\Phi)$. When we have calculated for
$\Phi(T)<\Phi_{0}(T)$ $F(\Phi)$ as well as $f(\Phi)$, i.e., in
compliance with both equations we also have calculated for
$\Phi(T)<\Phi_{0}(T)$ the function $P(\Phi)$. Therefore we can now
state that $P(\Phi)$ is also a polynomial function with positive
coefficients $\lambda_{n}$, which will fit for
$\Phi(T)<\Phi_{0}(T)$ the function $g(\Phi)$ to $G(\Phi)$.

\begin{figure}[htbp]
 \centering \leavevmode
\includegraphics[width=8.cm, height=5.cm]{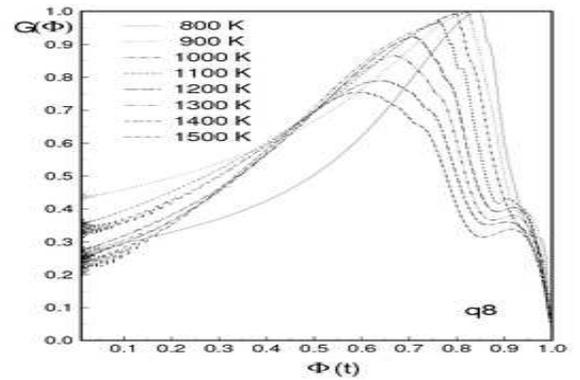}
 \caption{$G(\Phi)$ for wave vector $q_8$ according to Eq.(\ref{Gphi}).
\label{Gtptmd8}}
\end{figure}

\begin{figure}[htbp]
\centering \leavevmode
\includegraphics[width=8.cm, height=5.cm]{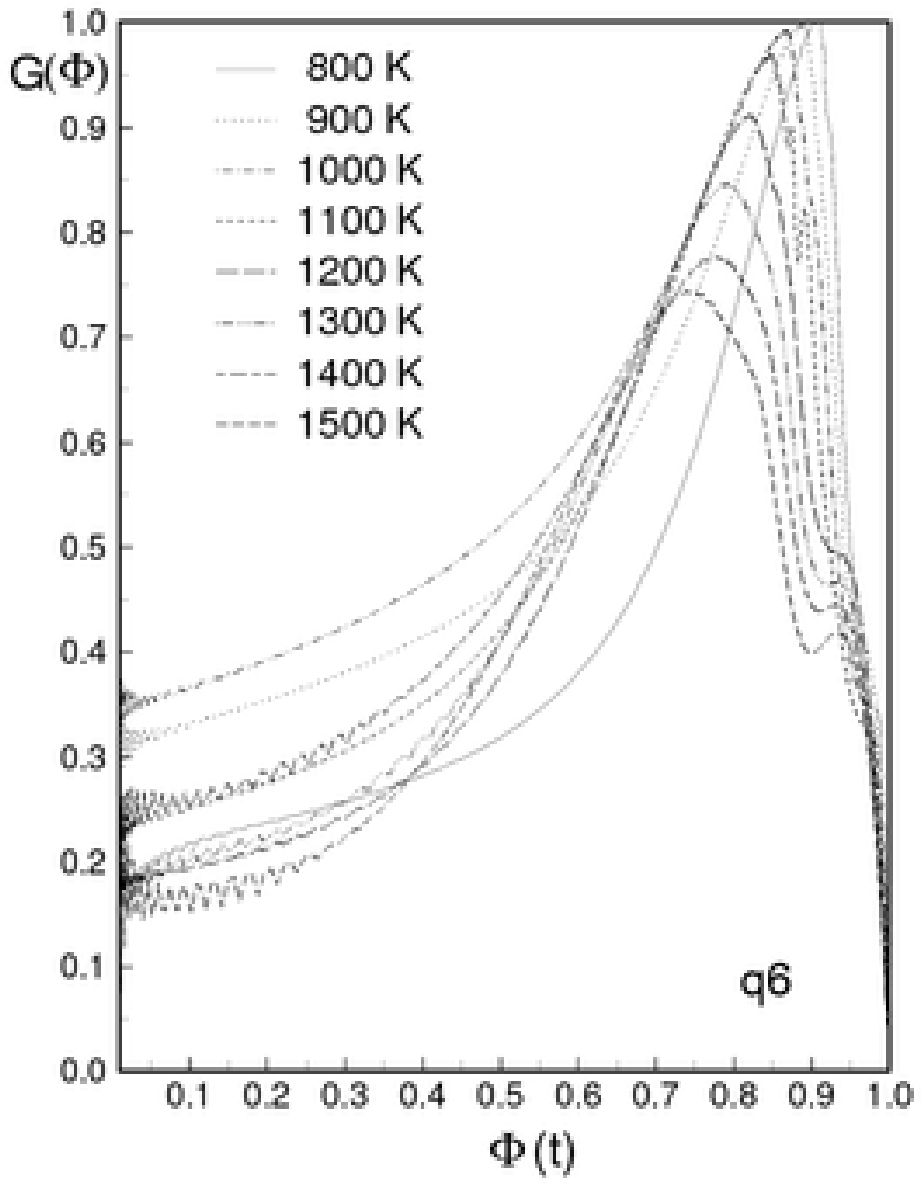}
\caption{The same as Fig.~\ref{Gtptmd8} but for wave vector
$q_{6}$. \label{Gtptmd6}}
\end{figure}

\begin{figure}[htbp]
 \centering \leavevmode
\includegraphics[width=8.cm, height=5.cm]{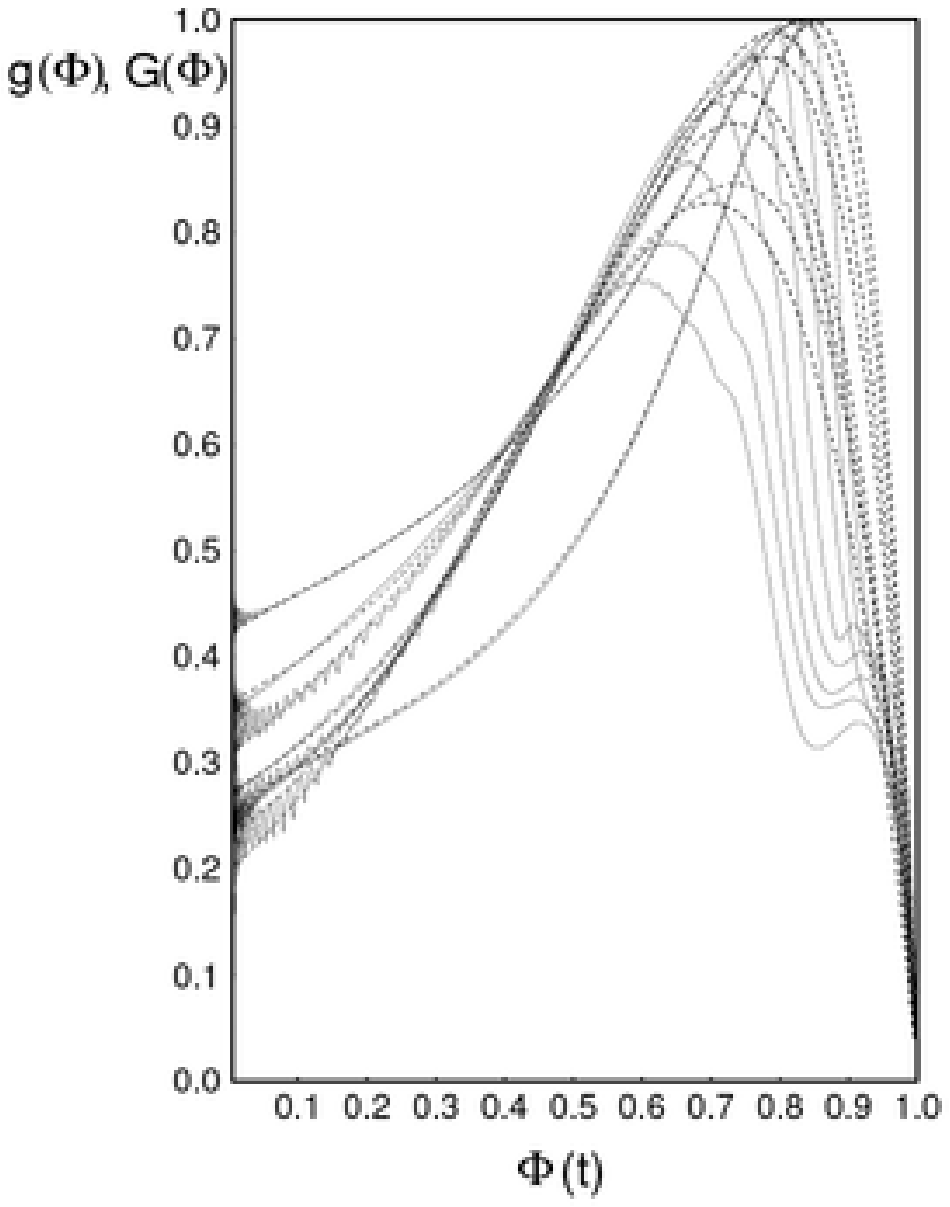}
\caption{ Fitting $g(\Phi)$ (dotted lines) to $G(\Phi)$ (full
lines) for wave vector $q_8$ according to Eqs.(\ref{gphi}) and
(\ref{Gphi}). \label{Anpgab1}}
\end{figure}

\begin{figure}[htbp]
\centering \leavevmode
\includegraphics[width=8.cm, height=6.cm]{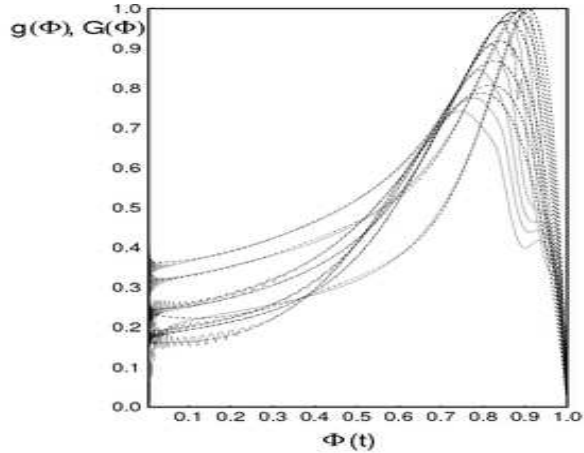}
\caption{ Fitting $g(\Phi)$ (dotted line) to $G(\Phi)$ (full
lines) for wave vector $q_6$ according to Eqs.(\ref{gphi}) and
(\ref{Gphi}). \label{Anpgab2}}
\end{figure}

In Figs.~\ref{Gtptmd8} and ~\ref{Gtptmd6} we present the
$G(\Phi)$-function. From both figures we can see that $G(\Phi)$
has a maximum at $\Phi_0(T)$. This maximum shifts with a
decreasing $q$-value to a higher value. We can now fit the
function $g(\Phi)$, which contains the polynomial function
$P(\Phi)$ with positive coefficients (see Tables.~\ref{table2} and
~\ref{table3}), to $G(\Phi)$. Our results is shown in Figs.
~\ref{Anpgab1} and ~\ref{Anpgab2}. From both figures we have for
each temperature  a maximum $g_m$ from $g(\Phi)$-function at
$\Phi=\Phi_m$. This maximum is smaller als 1 for $T\geq1100$ K (
$g_m<1 $ ), and for $T\leq 1000$ K nearly equal 1 ( $g_m \approx 1
$).

\begin{figure}[htbp]
\centering \leavevmode
\includegraphics[width=8.cm, height=5.cm]{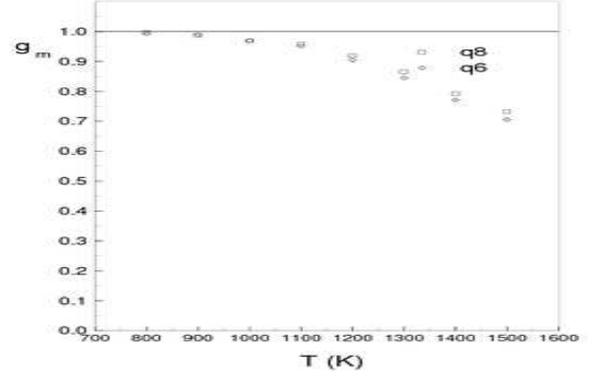}
\caption{$g_m$ as a
function of $T$. \label{Gtmax}}
\end{figure}

As mentioned in the theory section, $g_m$ describes a signature,
so that, if $g_m<1$, the system is in a ergodic state (in a liquid
state), and if $g_m>1$, it is in a non-ergodic state (in a
structural arrest). Our results show that the
Ni$_{20}$Zr$_{80}$-system is for $T\ge1100$ K in a liquid state
and for $T\leq1000$ K nearly in a structural arrest with a
correlation decay, which is as a result of a thermally activitated
atomic diffusion. In Fig.~\ref{Gtmax} we present a curve $g_m$ vs
$T$ for each $q$-value. From that Fig. one can see that our system
has a critical temperature $T_c$ between 1000 K and 1100 K. Here
we decide $T_c$ as about (1025 $\pm$ 25) K.

\subsection{Gaussian Approximation \label{Gsana}}

In following Figs. \ref{Pmsv98}, \ref{Pmsv108}, \ref{Pmsv118}, and
\ref{Pmsv128} we have calculated the self-part intermediate
scattering function $\Phi(q,t)$ by using Gaussian approximation,
also according to Eqs.(\ref{3.16.0}) and (\ref{3.17}). As input
data for both equations are the mean squared displacement (MSD)
and the velocity autocorrelations functions obtained from our
MD-Data. From figures we found that the gaussian approximation is
correct in the short-time regime, here also for $t<0.2$ ps.  For
$t>0.2$ ps the course of $\Phi(q,t)$ resulting from this
approximation deviates from both MD-results and
MCT-analysis-results. As mentioned in Subsection \ref{GS} we have
assumed that the DOS of phonons is equally approximated to the
spectral distribution of velocity autocorrelations function
(VACF). With this assumption we can state now that the
phonon-regime in our system is the regime taken place in a range
time smaller then 0.2 ps. This time is a nearly
temperature-independent time.

\begin{figure}[htbp]
 \centering \leavevmode
\includegraphics[width=8.cm, height=5.cm]{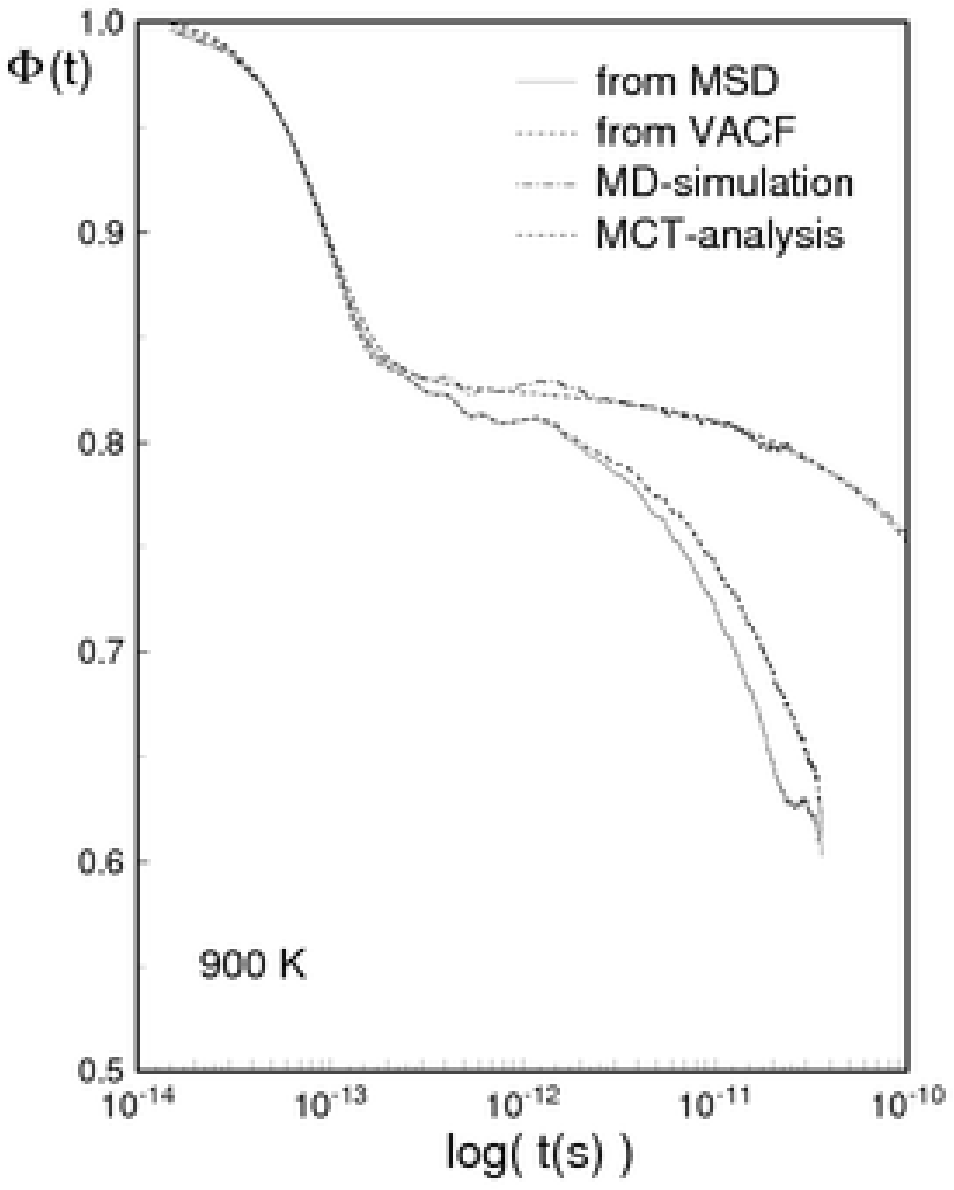}
\caption{ The self-part of the intermediate scattering function
for $T=$ 900 K, according to Eqs. (\ref{3.16.0})(line) and
(\ref{3.17})(dotted line), MD-result (dotted - dashed line),
MCT-analysis (dashed line). \label{Pmsv98}}
\end{figure}

\begin{figure}[htbp]
\centering \leavevmode
\includegraphics[width=8.cm, height=5.cm]{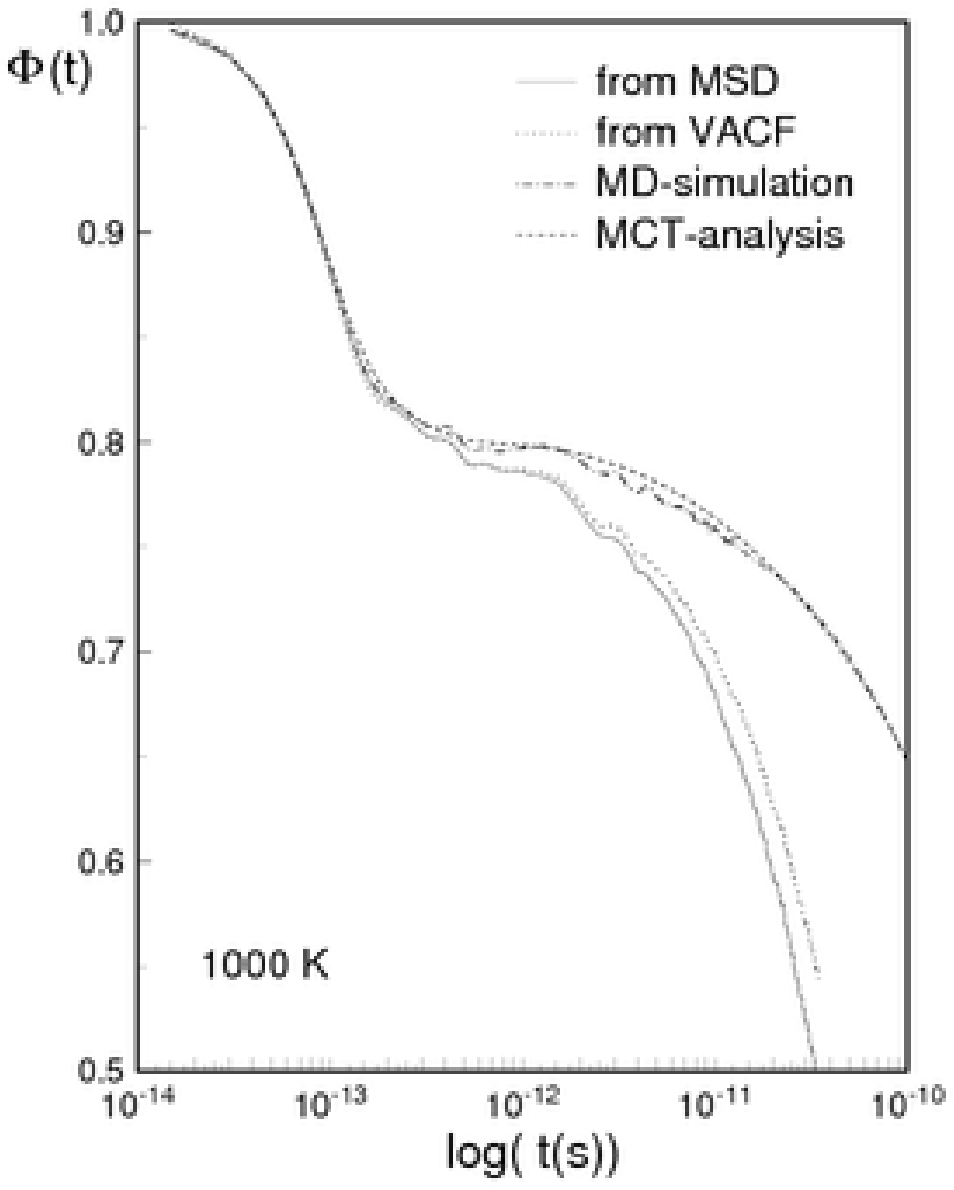}
\caption{ The same as Fig.\ref{Pmsv98} but for $T=$ 1000 K.
\label{Pmsv108}}
\end{figure}

\begin{figure}[htbp]
 \centering \leavevmode
 \includegraphics[width=8.cm, height=5.cm]{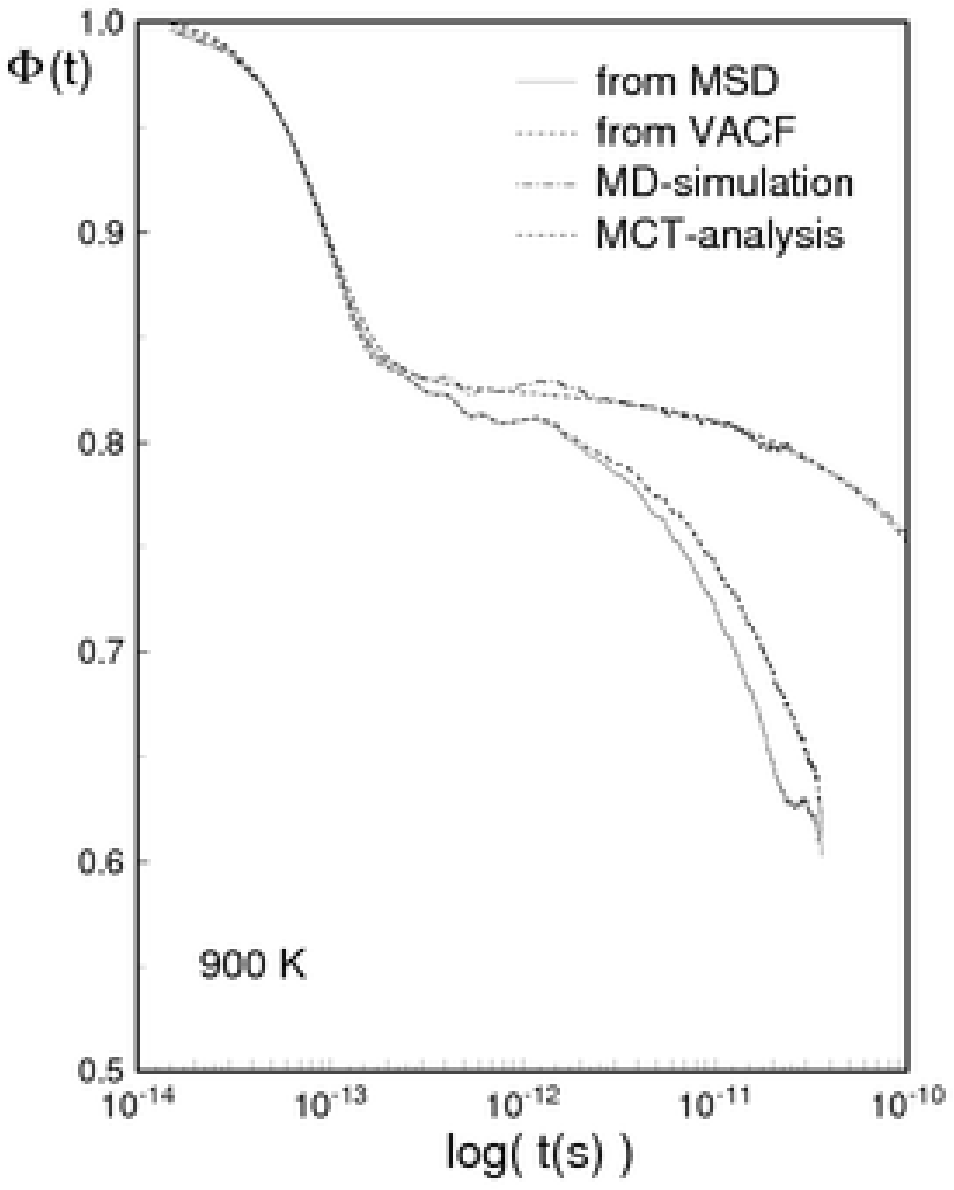}
 \caption{ The same as Fig.\ref{Pmsv98} but for $T=$ 1100 K.
\label{Pmsv118}}
\end{figure}

\begin{figure}[htbp]
\centering \leavevmode
\includegraphics[width=8.cm, height=5.cm]{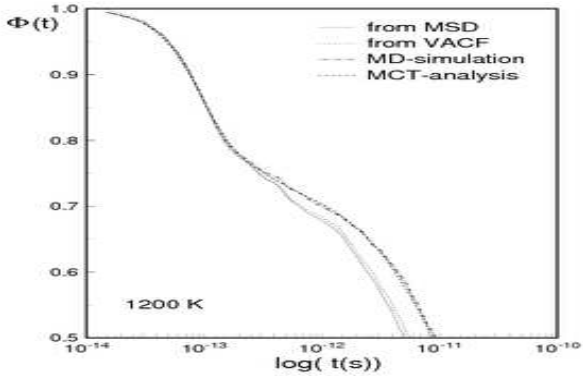}
\caption{ The same as Fig.\ref{Pmsv98} but for $T=$ 1200 K.
\label{Pmsv128}}
\end{figure}

To gain more our statement that this regime is as a result of the
phonon vibrations we borrow a important relation of the liquid
theory, namely, the relation between the spectral distribution of
VACF and the susceptibility of the self-part intermediate
scattering function $\Phi(q,t)$. According to theory that relation
is approximated for a small $q$-value (lim $q \rightarrow 0$) as
follows \cite{Boon,Hansen}
\begin{equation}
   \chi^{\prime\prime}(\omega)=\omega \Phi_c(\omega) \approx q^2 Z(\omega)/\omega
 \label{4.10}
\end{equation}

\begin{figure}[htbp]
\centering \leavevmode
\includegraphics[width=8.cm, height=5.cm]{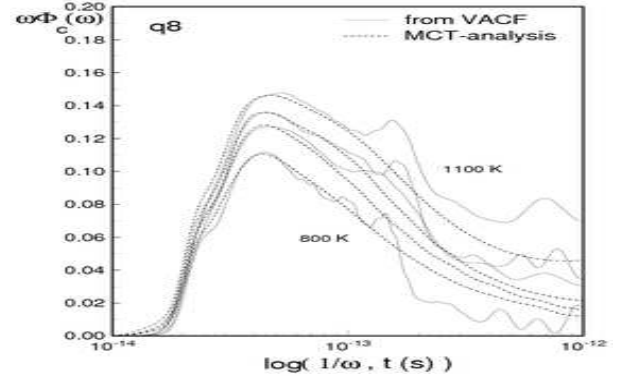}
\caption{The susceptibility obtained from the spectral
distribution of VACF through Eq.(\ref{4.10}) (lines) and from the
MCT-analysis (dashed lines) \label{Spgab}}
\end{figure}

Fig.\ref{Spgab} presents our calculating results of
Eq.(\ref{4.10}) with an assumption that our $q_8$-value is small
enough. The calculating results show a fast nearly good agreement
with that from the MCT-analysis (or a directly Fourier-tranformed
$\Phi(q,t)$).

\begin{figure}[htbp]
 \centering \leavevmode
\includegraphics[width=8.cm, height=5.cm]{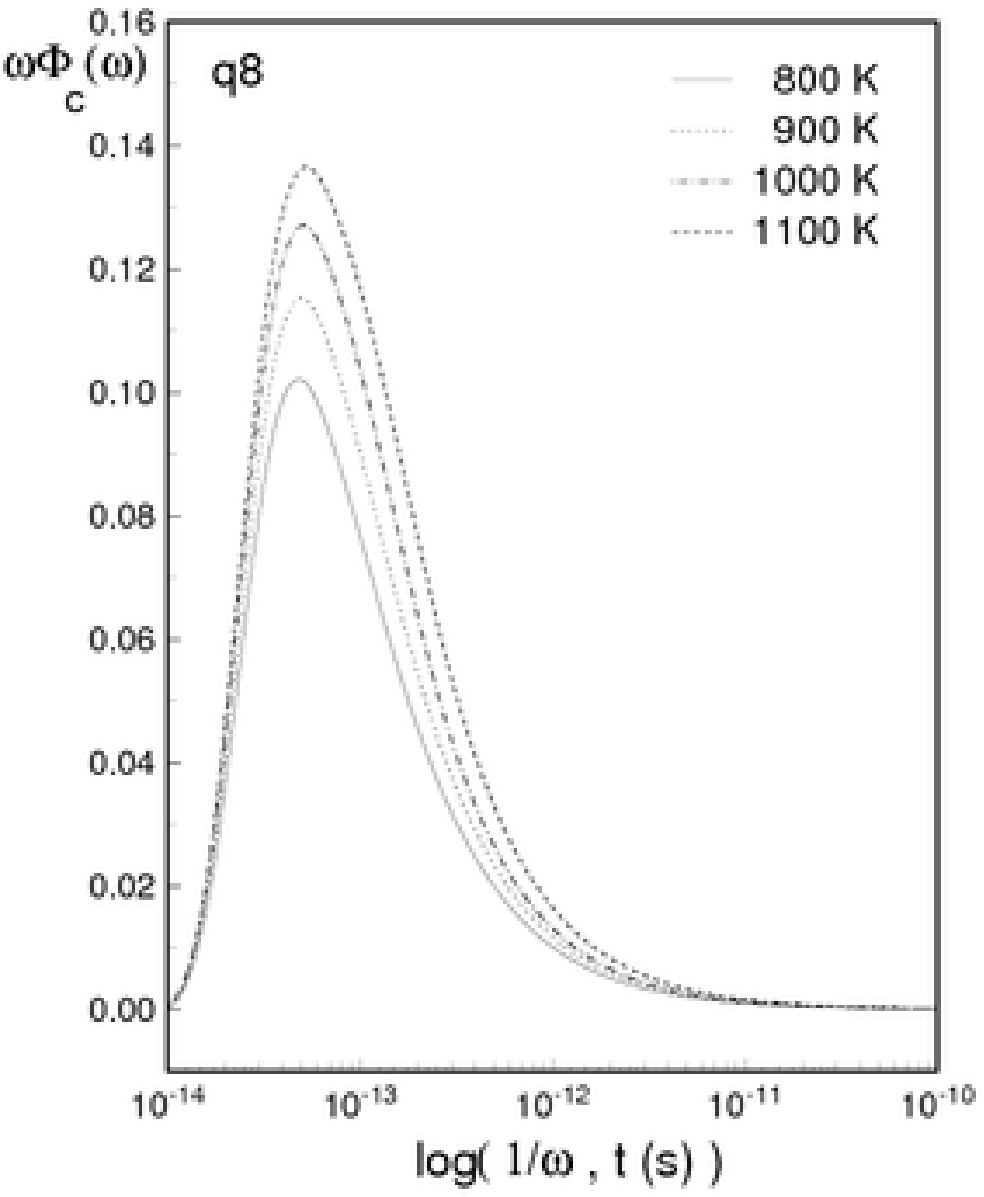}
 \caption{ The susceptibility obtained before superposition for wave vector $q_8$.
\label{Krvorsk8}}
\end{figure}

\begin{figure}[htbp]
\centering \leavevmode
\includegraphics[width=8.cm, height=5.cm]{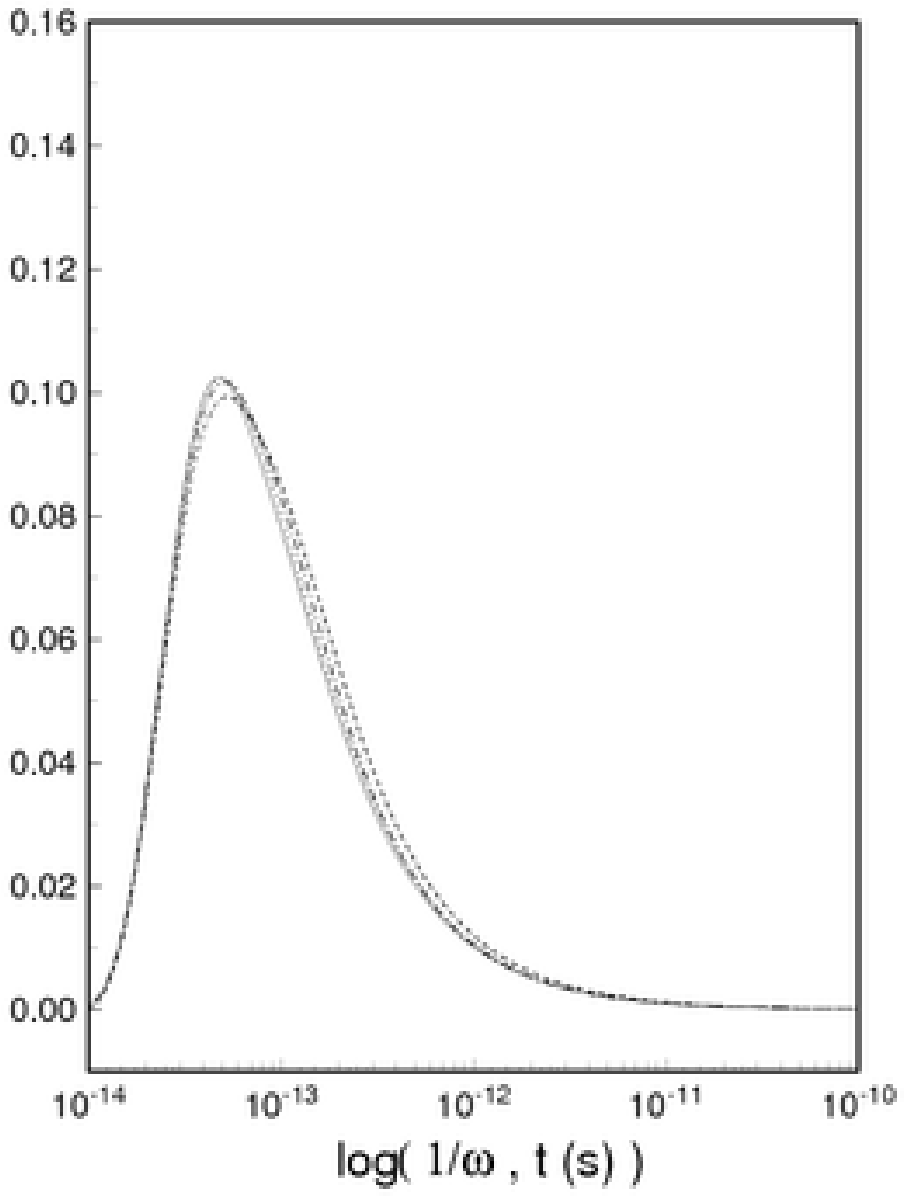}
\caption{ The susceptibility obtained after superposition.
\label{Krsk8}}
\end{figure}
\begin{figure}[htbp]
 \centering \leavevmode
 \includegraphics[width=8.cm, height=5.cm]{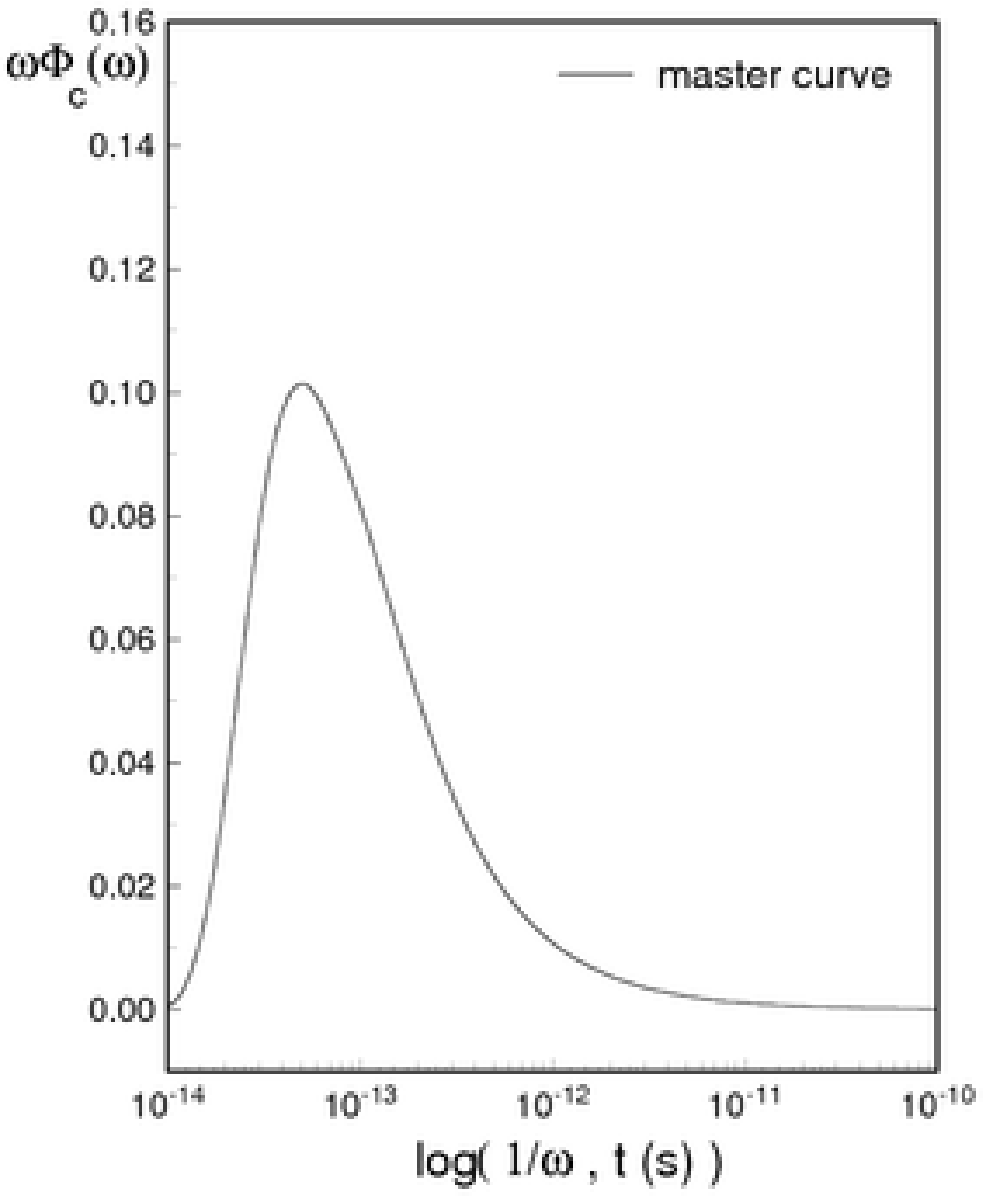}
 \caption{ The master curve of the susceptibility for $q_8$ according to Eq.(\ref{3.14}).
\label{Mst8}}
\end{figure}
\begin{figure}[htbp]
\centering \leavevmode
\includegraphics[width=8.cm, height=5.cm]{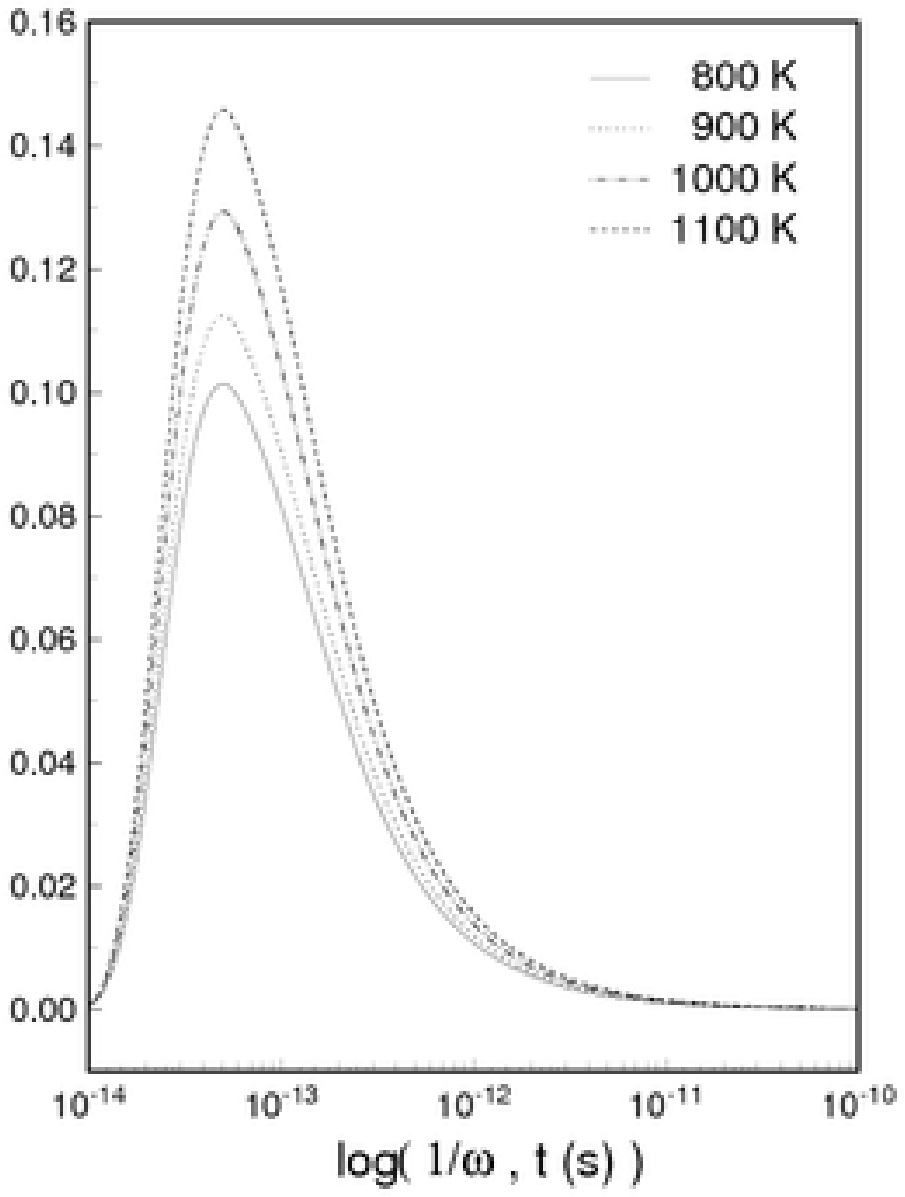}
\caption{ The new susceptibility after scaling for $q_8$ according
to Eq.(\ref{3.13}). \label{Krnask8}}
\end{figure}

\begin{figure}[htbp]
 \centering \leavevmode
\includegraphics[width=8.cm, height=5.cm]{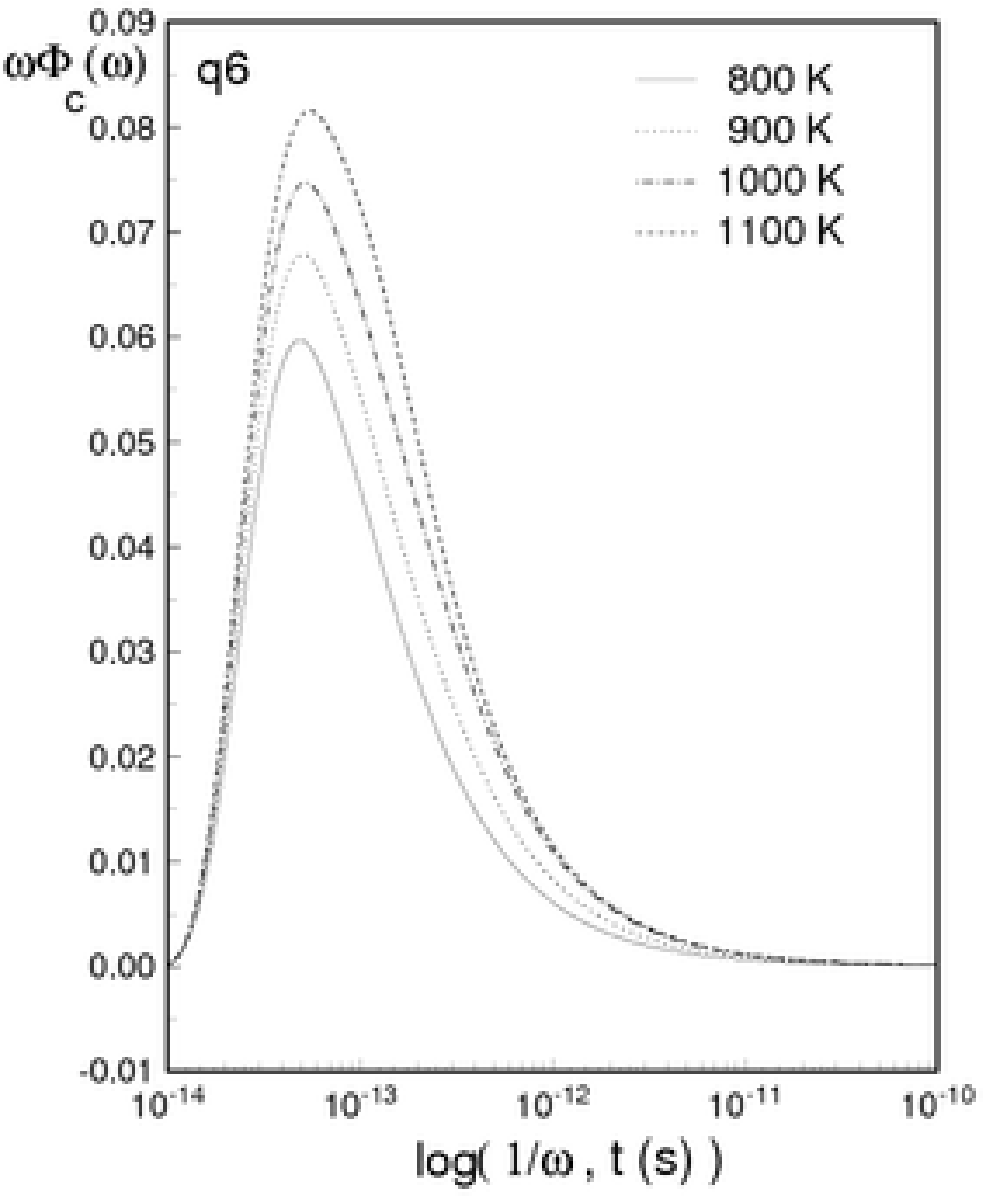}
 \caption{The susceptibility obtained before superposition for wave vector $q_6$.
\label{Krvorsk6}}
\end{figure}

\begin{figure}[htbp]
\centering \leavevmode
\includegraphics[width=8.cm, height=5.cm]{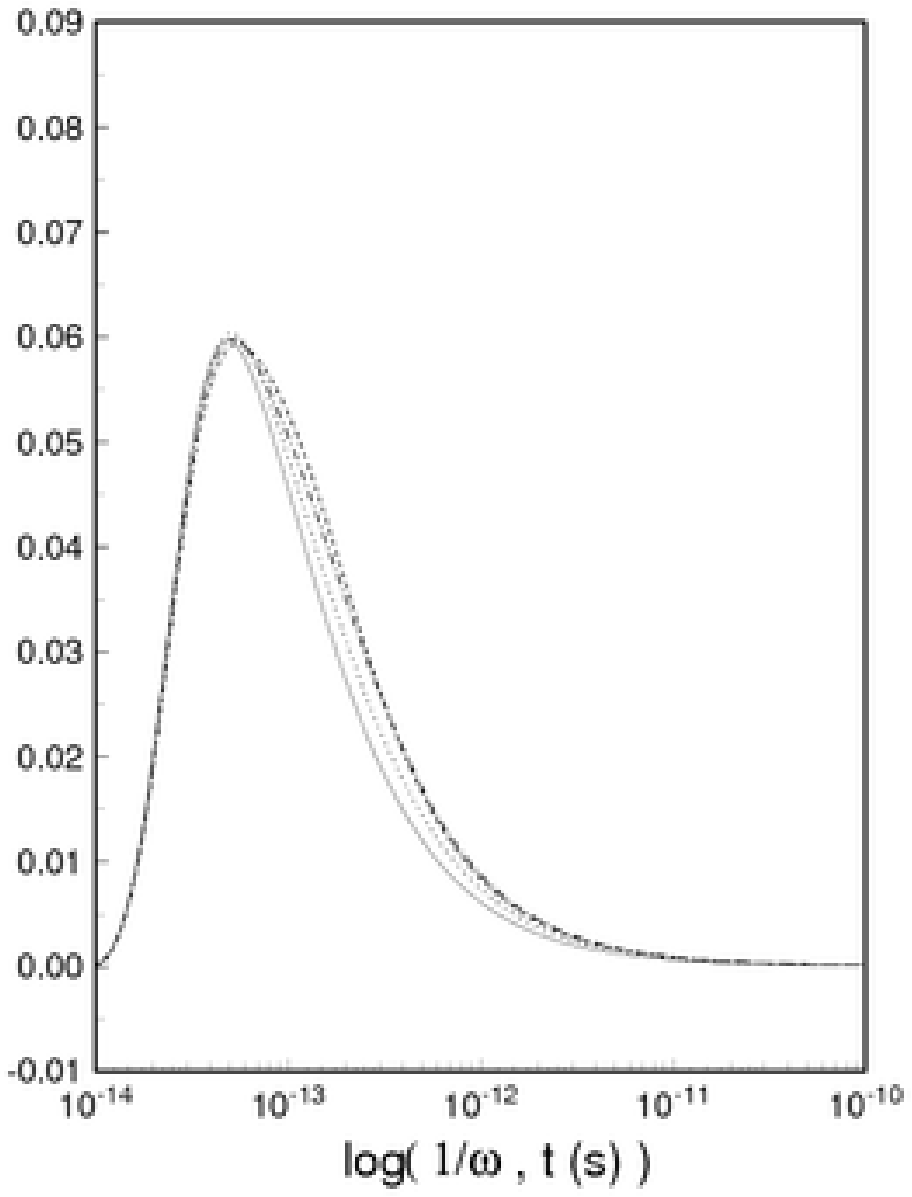}
\caption{The susceptibility obtained after superposition.
\label{Krsk6}}
\end{figure}

\begin{figure}[htbp]
 \centering \leavevmode
\includegraphics[width=8.cm, height=5.cm]{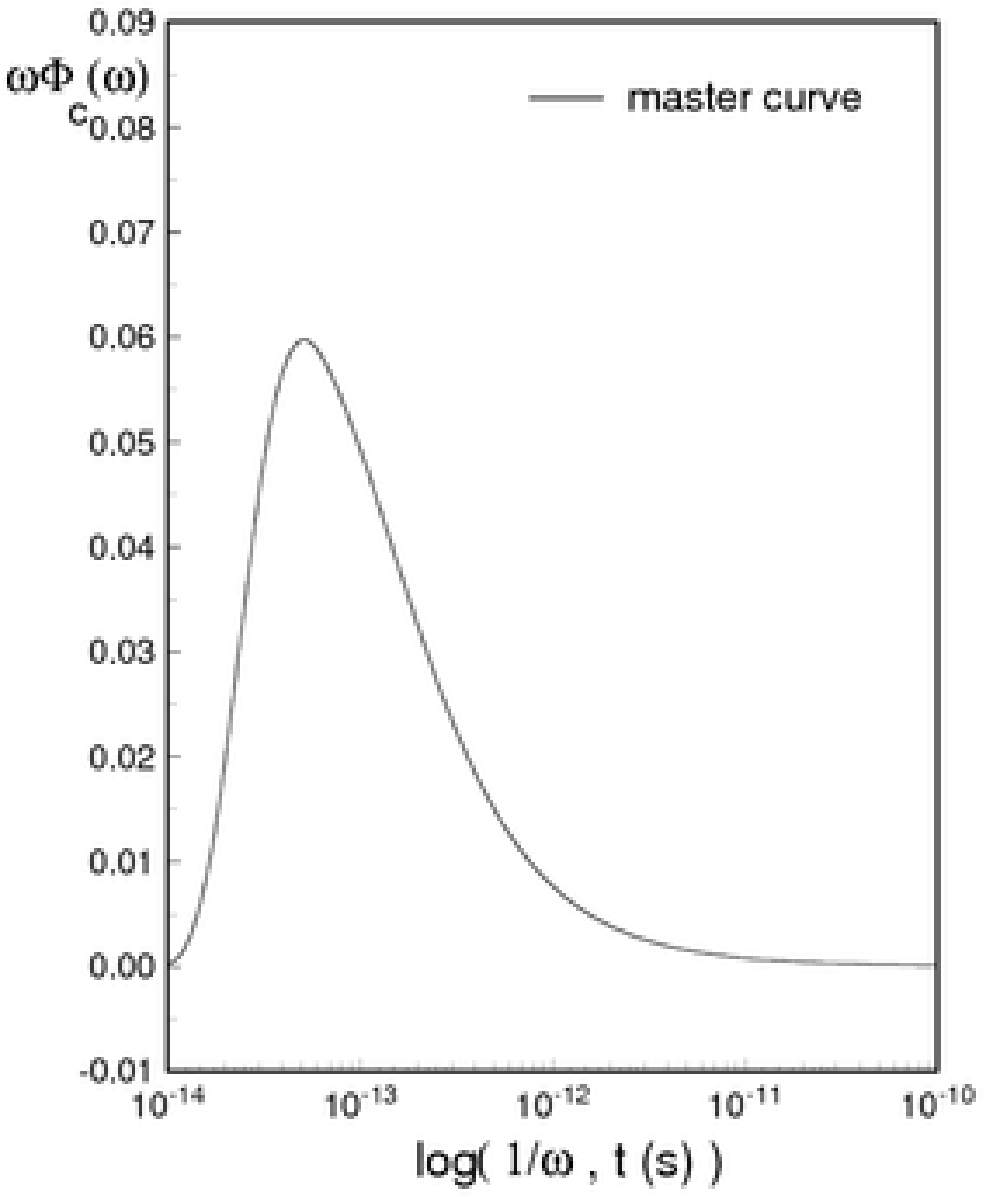}
 \caption{The master curve of the susceptibility for $q_6$ according to Eq.(\ref{3.14}).
\label{Mst6}}
\end{figure}
\begin{figure}[htbp]
\centering \leavevmode
\includegraphics[width=8.cm, height=5.cm]{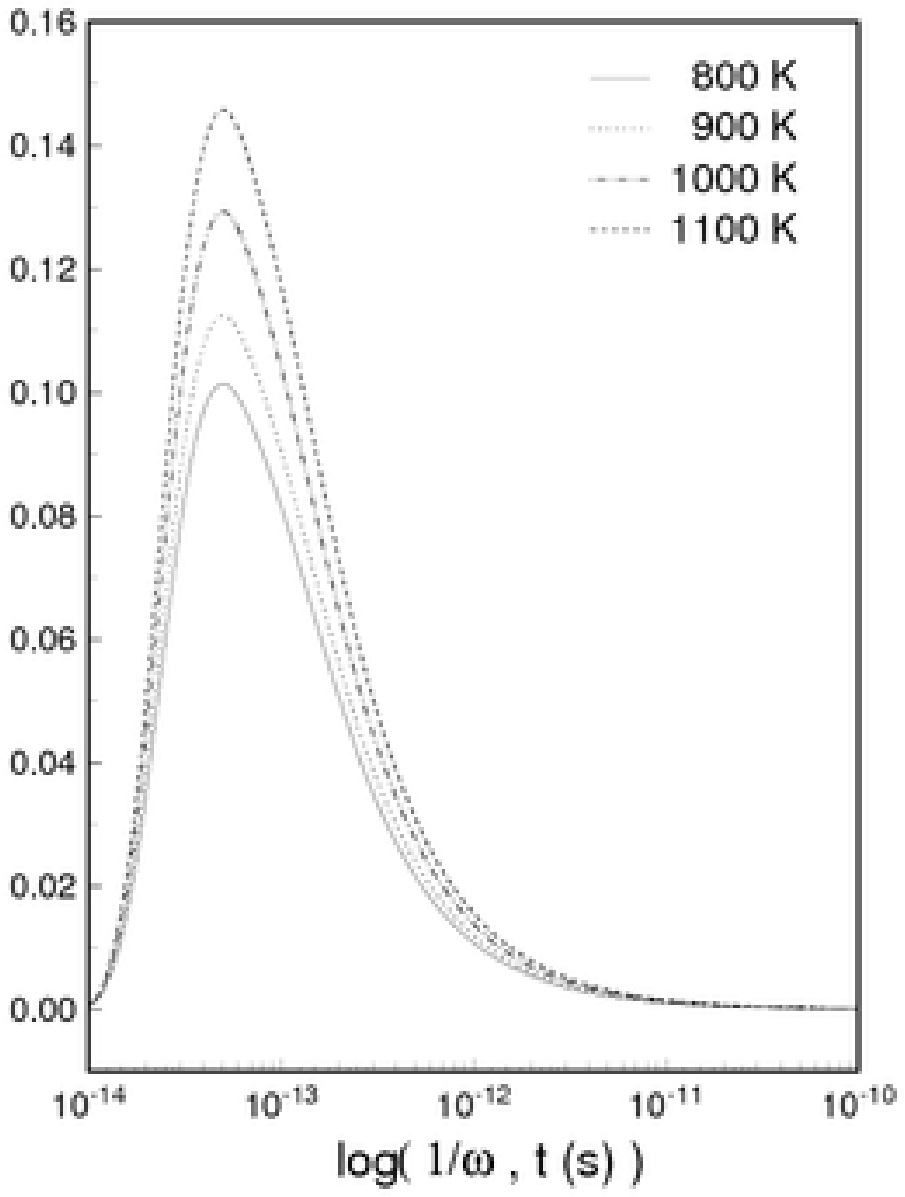}
\caption{ The new susceptibility after scaling for $q_6$ according
to Eq.(\ref{3.13}). \label{Krnask6}}
\end{figure}

Actually we can do a inverse Fourier-transformation of the
susceptibility obtained from Eq.(\ref{4.10}) to find back the
self-part intermediate scattering function $\Phi(q,t)$, but from
the result shown in Fig.\ref{Spgab} we do not do it, because
Eq.(\ref{4.10}) is a approximated relation that is really correct
if the wave vector $q$ is very small (lim $q \rightarrow 0$ or, as
one calls it, $q$ is in a hydrodynamics limit), and then if we
attempt to back Fourier transform of its susceptibility, surely we
can not find a better $\Phi(q,t)$ as it found through
Eq.(\ref{3.17}).

Further discussion about susceptibility we attempt to separate the
susceptibility of this phonon-regime from the full susceptibility.
As mentioned in Sec. \ref{THEO} we assume that this separation is
really to find a purely susceptibility of this regime. Our result
for two wave vector is shown in Figs. \ref{Krvorsk8} and
\ref{Krvorsk6}. We scale now these susceptibilities, the result is
shown in Figs.\ref{Krsk8} and \ref{Krsk6}, here we have chosen as
a arbitrary normalization of temperature, $T_0$, 800 K, and we
average these scaled susceptibilities through Eq.({3.14}) to
obtain the master curve of the susceptibility in this regime(see
Figs.\ref{Mst8} and \ref{Mst6}). By using Eq.(\ref{3.13}) we
obtain the new susceptibility of this regime (see Figs.
\ref{Krnask8} and \ref{Krnask6}).

\begin{figure}[htbp]
\centering \leavevmode
\includegraphics[width=8.cm, height=5.cm]{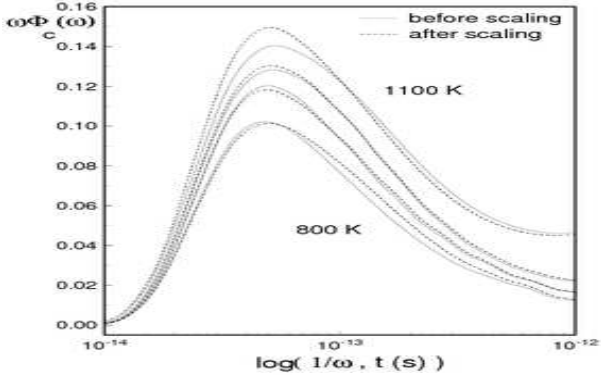}
\caption{ The full susceptibility before (dashed lines) and after
scaling(full lines) in a short-time regime for wave vector $q_8$.
\label{Mdkskl8}}
\end{figure}

\begin{figure}[htbp]
\centering \leavevmode
\includegraphics[width=8.cm, height=5.cm]{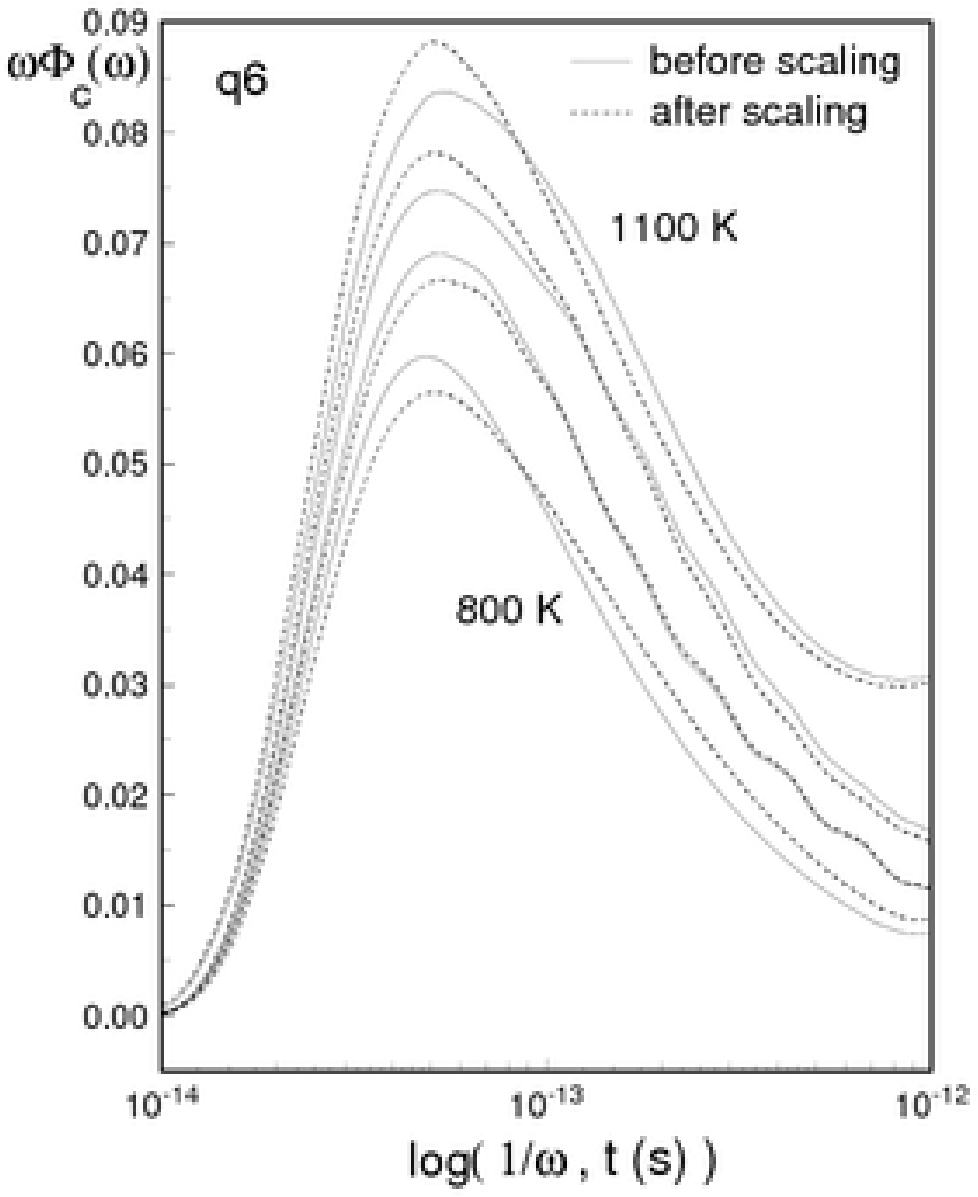}
\caption{ The full susceptibility before (dashed lines) and after
scaling(full lines) in a short-time regime for wave vector $q_6$.
\label{Mdkskl6}}
\end{figure}

Adding the new susceptibility as shown in Figs.\ref{Krnask8} and
\ref{Krnask6} with the susceptibilities from other regimes, then
we obtain the full susceptibility in the short-time regime as
shown in Figs. \ref{Mdkskl8} and \ref{Mdkskl6}. From both figures
we found that the the peaks of the new full susceptibility for
temperatures  1000 K and 1100 K are higher then those old ones,
but that peaks for temperatures 800 K and 900 K are lower then
those old ones.

\begin{figure}[htbp]
 \centering \leavevmode
\includegraphics[width=8.cm, height=5.cm]{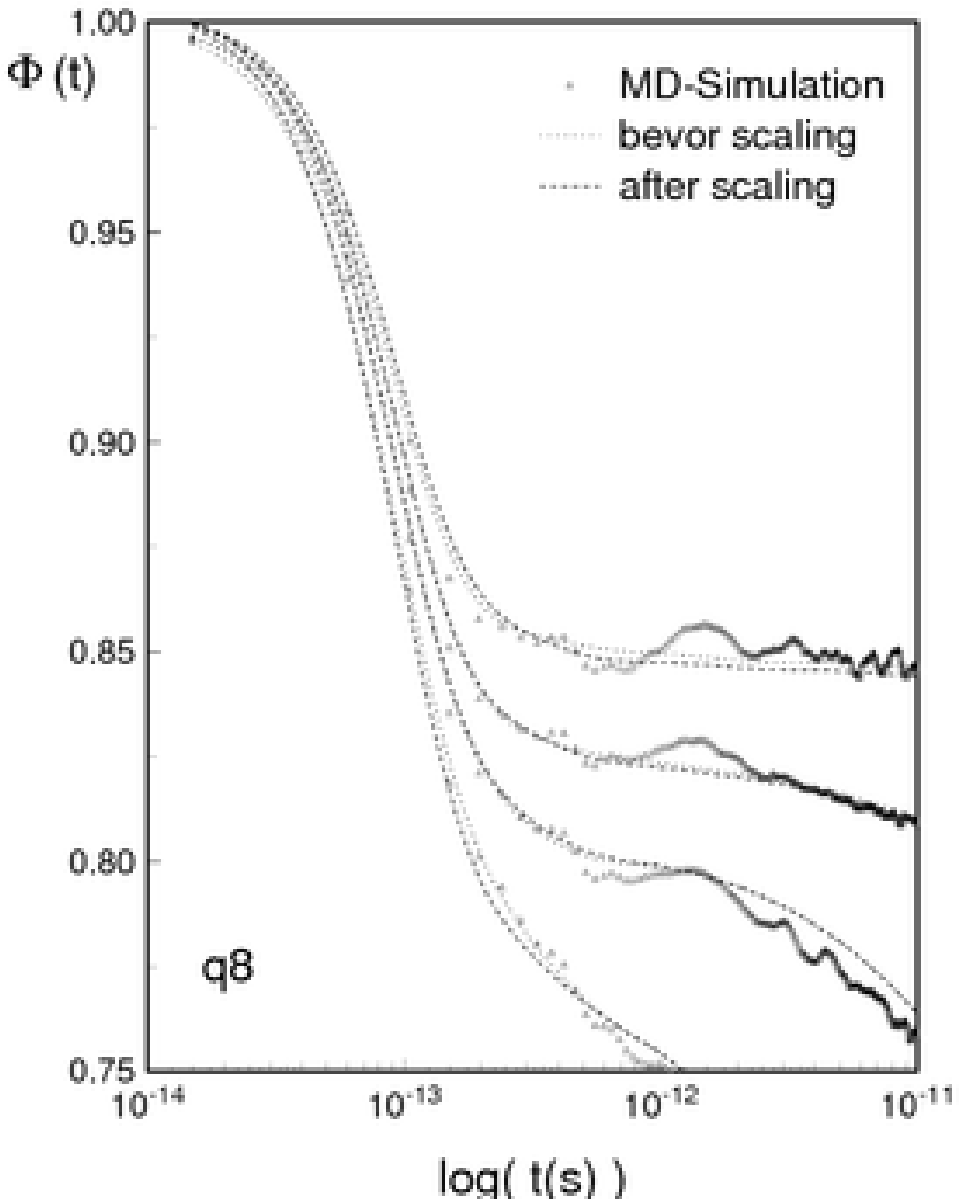}
\caption{ The self-part of the intermediate scattering function in
a short-time regime, MD-results (Symbol), before (dotted line) and
after scaling (dashed-line) for wave vector $q_8$.
\label{Pmmctsk8}}
\end{figure}

\begin{figure}[htbp]
\centering \leavevmode
\includegraphics[width=8.cm, height=5.cm]{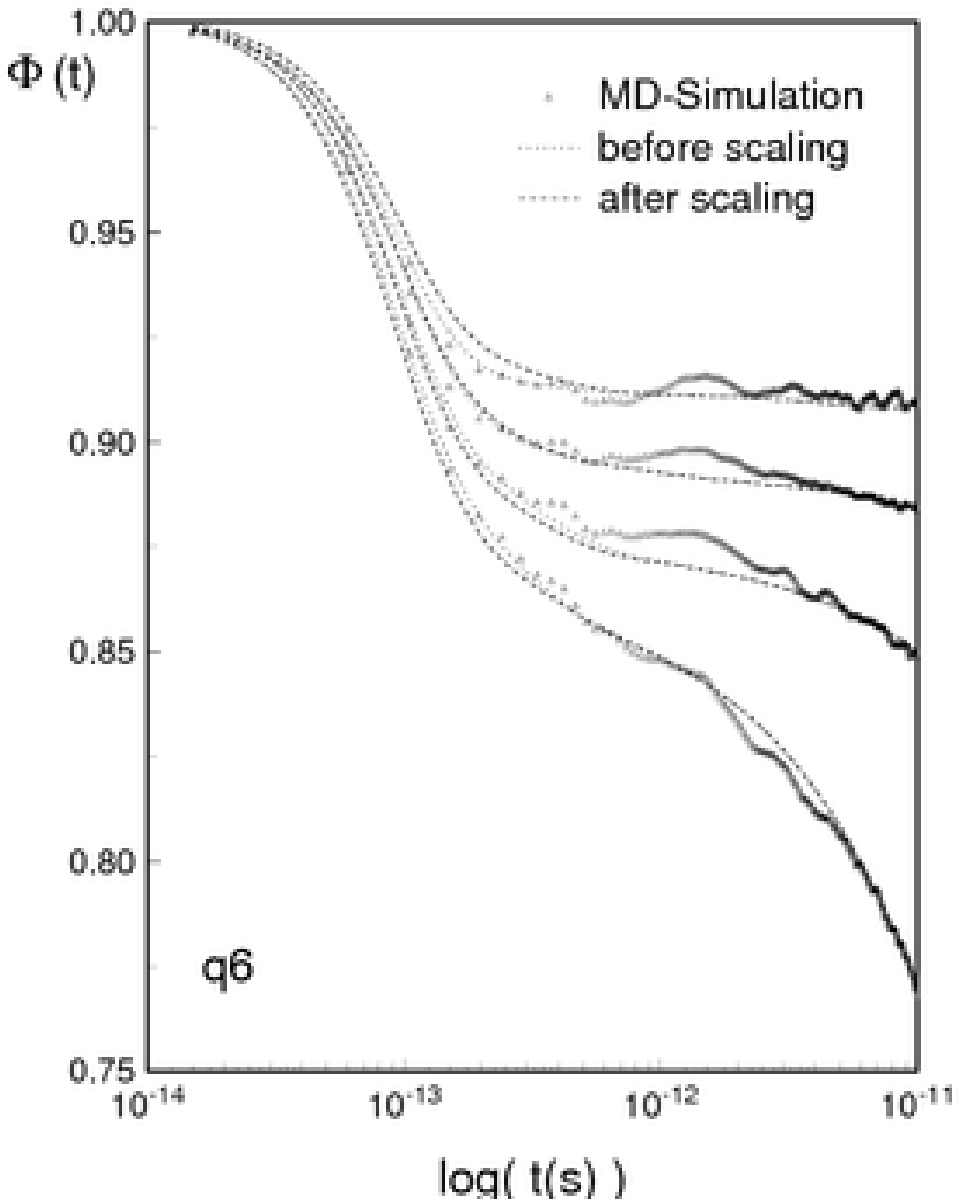}
\caption{ The self-part of the intermediate scattering function in
a short-time regime, MD-results (Symbol), before (dotted line) and
after scaling (dashed-line) for wave vector $q_6$.
\label{Pmmctsk6}}
\end{figure}

\begin{figure}[htbp]
 \centering \leavevmode
\includegraphics[width=8.cm, height=5.cm]{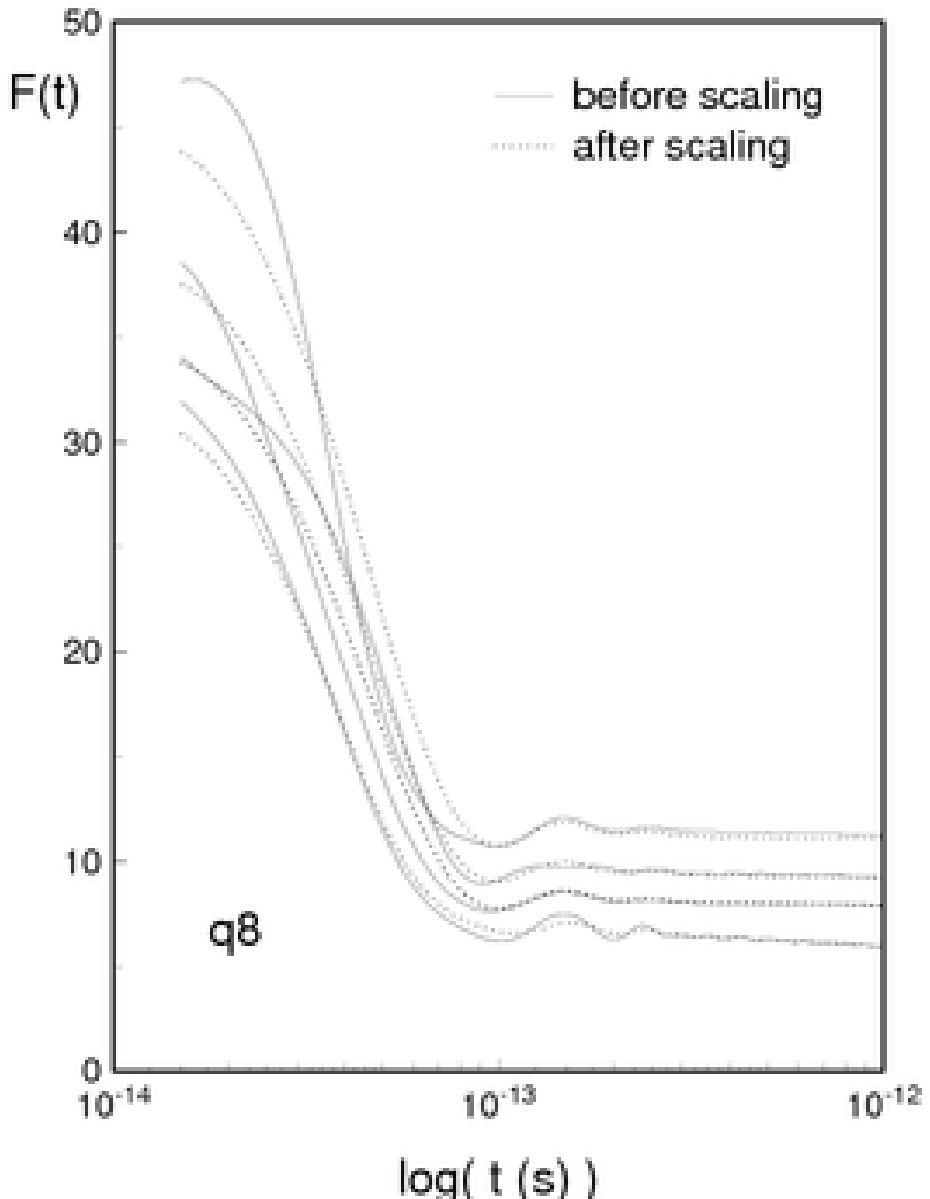}
\caption{ The kernel $F(t)$ in the short-time regime before (full
lines) and after scaling (dotted lines) for wave vector $q_8$.
\label{Ft8sklvg}}
\end{figure}

\begin{figure}[htbp]
\centering \leavevmode
\includegraphics[width=8.cm, height=5.cm]{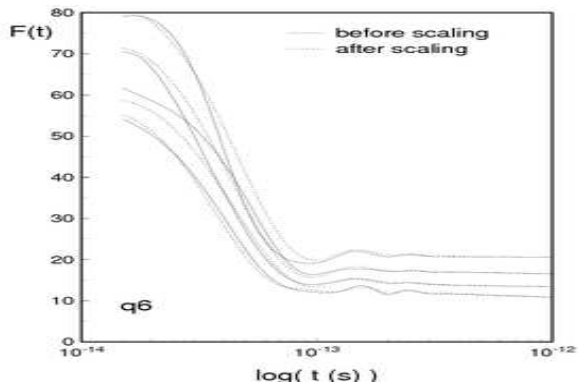}
\caption{ The kernel $F(t)$ in the short-time regime before (full
lines) and after scaling (dotted lines) for wave vector $q_6$.
\label{Ft6sklvg}}
\end{figure}



Taking back Fourier transformation of $\Phi_c(\omega)$ of the new
full susceptibility in the short-time regime, then we obtain the
new $\Phi(q,t)$ as shown in Figs. \ref{Pmmctsk8} and
\ref{Pmmctsk6}. The deviation of the new $\Phi(q,t)$ from the old
one for each temperature of interest is taken place at a time
range that is smaller then 0.5 ps.

It is interesting to calculate the new kernel $F(t)$ and the new
$G(\Phi)$-function in this short-time regime. Our results for the
kernel $F(t)$ are shown in Figs. \ref{Ft8sklvg} and
\ref{Ft6sklvg}. The new kernel $F(t)$ for each temperatures of
interest show more regular then the old one. The deviation of the
new kernel from the old one is taken place at a time range that is
smaller then 0.5 ps, also this time range is the same as it found
in the case of $\Phi(q,t)$. It is clear that $\Phi(q,t)$ in this
regime depends strongly on the kernel $F(t)$.

\section{CONCLUDING REMARKS}
Our system behaves as predicted by the schematic MCT-Modell, in
the sense that the self-part intermediate scattering function
$\Phi(q,t)$ shows at lower temperature three step of the
structural relaxation. The behavior of the self-part intermediate
scattering function $\Phi(q,t)$ agrees well for the long time with
the prediction of the extended schematic MCT-Model in the sense
that the self-part intermediate scattering function $\Phi(q,t) $
always decays to zero after long times as a result of thermally
activated atomic diffusion.

The behavior of the memory kernel $F(t)$ agrees with the
prediction of the extended MCT-Model in the sense that the
behavior of $F(t)$ below $\Phi_0(t)$ corresponds to a polynomial
function that has non-negative coefficients. This behavior depends
strongly on the wave vector $q$.

As in I, to determine approximately the critical temperature $T_c$
we have used the maximum $g_m$ of the characteristic function
$g(\Phi)$. We obtain $T_c\approx(1025\pm25)$ K for our system.

Through the temperature superposition method in a short times
regimes our results show that the vibration-phonon regime is
really as a result of the harmonic phonon approximation, and then
this regime can be good described by a gaussian approximation. In
the case of a gaussian approximation, which follows from the
liquid state theory, the courses of the self-part intermediate
scattering function $\Phi(q,t)$ resulted from both VACF and MSD
shows at time, $t\approx0.5$ ps a deviation from the proper
courses obtained from MD-simulations and MCT-analysis

\acknowledgements A.B.M. gratefully  acknowledges financial
support of the DAAD during the post-doctoral program.

\end{document}